\documentclass[amsfonts,amssymb]{article}

\usepackage{graphicx}
\pdfoutput=1
\usepackage{subcaption}
\usepackage[titletoc]{appendix}
\usepackage{color}

\renewcommand{\theequation}{\arabic{equation}}
\newcommand{\be}{\begin{equation}}
\newcommand{\ee}{\end{equation}}
\newcommand{\bea}{\begin{array}}
\newcommand{\ea}{\end{array}}
\newcommand{\beqa}{\begin{eqnarray}}
\newcommand{\eeqa}{\end{eqnarray}}

\newcommand{\eean}{\end{eqnarray*}}

\def\up#1{\leavevmode \raise.16ex\hbox{#1}}

\newcommand{\gapproxeq}{\lower
 .7ex\hbox{$\;\stackrel{\textstyle >}{\sim}\;$}}
\newcommand{\lapproxeq}{\lower .7ex\hbox{$\;\stackrel
{\textstyle <}{\sim}\;$}}

\renewcommand{\theequation}{\arabic{section}.\arabic{equation}}
\newcounter{appendice}

\def\thebibliography#1{{\bf REFERENCES\markboth
 {REFERENCES}{REFERENCES}}\list
 {[\arabic{enumi}]}{\settowidth\labelwidth{[#1]}\leftmargin\labelwidth
 \advance\leftmargin\labelsep
 \usecounter{enumi}}
 \def\newblock{\hskip .11em plus .33em minus -.07em}
 \sloppy
 \sfcode`\.=1000\relax}

\def\BI{{\rm 1\!l}}

\begin{document}

\centerline{ \LARGE  Matrix Model Approach to Cosmology }

\vskip 2cm
\centerline{A. Chaney\footnote{adchaney@crimson.ua.edu}, Lei Lu\footnote{llv1@crimson.ua.edu} and A. Stern\footnote{astern@ua.edu} }
\vskip 1cm
\begin{center}
{ Department of Physics, University of Alabama,\\ Tuscaloosa,
Alabama 35487, USA\\}
\end{center}
\vskip 2cm
\vspace*{5mm}
\normalsize
\centerline{\bf ABSTRACT}

We perform a  systematic search for rotationally invariant cosmological solutions to matrix models, or more specifically the bosonic sector of Lorentzian  IKKT-type  matrix models, in dimensions $d$ less than ten, specifically $d=3$ and $d=5$.   After taking a continuum (or commutative) limit they yield $d-1$ dimensional space-time surfaces, with an attached Poisson structure,  which can be associated with closed, open or static cosmologies. For  $d=3$, we obtain recursion relations from which it is possible to generate rotationally invariant matrix solutions which yield open universes in the continuum limit.   Specific examples of matrix solutions have also been found which are associated with closed and static two-dimensional space-times in the continuum limit. The solutions  provide for  a matrix resolution of cosmological singularities.  The commutative limit reveals other desirable features, such as a solution describing a smooth transition from an initial inflation to a noninflationary era. Many of the $d=3$ solutions  have analogues in higher dimensions.  The case of  $d=5$, in particular, has the potential for yielding realistic four-dimensional cosmologies in the continuum limit. We find four-dimensional  de Sitter $dS^4$ or anti-de Sitter $AdS^4$ solutions  when a totally antisymmetric term is included in the matrix action. A nontrivial  Poisson structure is attached to these manifolds which represents the lowest order effect of noncommutativity.  For the case of $AdS^4$, we find  one particlular limit where the lowest order noncommutativity vanishes at the boundary, but not in the interior. 
\bigskip
\bigskip

\newpage
\tableofcontents

\section{Introduction}

  Space-time geometry,   field theory and  gravity have been shown to dynamically  emerge from matrix models.\cite{Aoki:1998vn},\cite{Steinacker:2010rh},\cite{Berenstein:2008eg},\cite{DelgadilloBlando:2008vi},\cite{Craps:2010cn},\cite{O'Connor:2011zz},\cite{Ydri:2015zsa} Moreover,   black hole  and cosmological space-times  were recovered in certain limits of matrix model solutions.\cite{Alvarez:1997fy},\cite{Freedman:2004xg},\cite{Craps:2005wd},\cite{Erdmenger:2007xs},\cite{Klammer:2009ku},\cite{Blaschke:2010ye},\cite{Kim:2011ts} Matrix models were introduced by  Ishibashi, Kawai, Kitazawa and  Tsuchiya (IKKT)\cite{Ishibashi:1996xs} and Banks, Fischler, Shenker, and Susskind (BFSS)\cite{Banks:1996vh} in order  to represent nonperturbative aspects of string theory.   Their application to black holes and cosmology  holds the promise of including such nonperturbative  effects.  In particular, it has the potential of resolving the singularities of general relativity.\cite{Craps:2011sp}

 The application of matrix models to cosmology first  appeared in a work by \'Alvarez and Meessen,\cite{Alvarez:1997fy} where a Newtonian type cosmology resulted from  the BFSS model.   Alternatively, because  time and spatial coordinates are treated in the same manner in the IKKT model,   a generally covariant theory  can appear from this model in the continuum limit. Much progress has been made by evaluating the quantum partition function for the IKKT model in Euclidean and Minkowski space.\cite{Aoki:1998vn},\cite{Berenstein:2008eg},\cite{DelgadilloBlando:2008vi},\cite{Craps:2010cn},\cite{O'Connor:2011zz},\cite{Kim:2011ts}   In \cite{Kim:2011ts} it was the  demonstration that the rotational invariance of nine spatial dimensions in the IKKT matrix model is spontaneously broken to $SO(3)$ acting with the defining representation  on three spatial dimensions.
This gives  justification for studying a simpler version of the IKKT model written in less than  ten dimensions, with the possible addition of terms in the action which preserve the symmetries of the lower dimensional space-time. In addition, examinations of classical, as well as quantum, aspects of matrix models have yielded interesting features related to  black holes and cosmology.  Solutions resembling expanding universes were obtained in \cite{Freedman:2004xg},\cite{Klammer:2009ku},\cite{Kim:2011ts}.  In \cite{Chaney:2013vha} a lower dimensional   matrix model  was written down from which it was possible to recover the usual BTZ black hole entropy formula. 
  In \cite{Chaney:2015mfa}  it was shown how space-time singularities can be resolved in a low-dimensional  matrix model.  There we found classical solutions which were the Lorentzian analogues of  fuzzy spheres.  They are expressed in terms of $N\times N$ matrices, where time and space were associated with discrete spectra.   The continuous, or  commutative, limit corresponds to  $N\rightarrow\infty$, and singularities on an otherwise smooth manifold  appeared upon taking this limit.  The manifold describes a closed   two dimensional cosmological space-time and the singularities resemble cosmic  singularities.

In this article we examine   the bosonic sector  of Lorentzian matrix models in less than ten dimensions, specifically,  dimensions $d=3$ and $5$.  
The matrix models are  of  the IKKT-type, as they are obtained by a reduction of a Yang-Mills theory to a zero-dimensional domain.  Terms are included in the action which respect the Lorentz symmetry of the theory.   One such term, which can be written down for any odd $d$, is totally antisymmetric in space-time indices and is analogous to a topological term.  We shall  search for `rotationally invariant' matrix solutions to the classical equations.  In the commutative limit, rotationally invariant configurations are associated with $d-1$ space-time dimensional  manifolds. Rotational invariant matrices in $2+1$ space-time are easy to define.  The dynamical degrees of freedom in this case are contained in three infinite-dimensional Hermitean matrices $Y^\mu$, $\mu=0,1,2$, with $0$ being  the time index and $1$ and $2$ being spatial indices.
We can take rotationally invariant matrix configurations for $(Y^0,Y^1,Y^2)$ to be those  satisfying 
 \be [Y_+Y_-,Y^0]= 0\label{cmxpxmwx0}\;,\ee
where
\be Y_\pm=Y^1\pm iY^2 \label{yplsy-}\ee
The commutative limit of a matrix model  is defined in analogous fashion to the classical limit  of a quantum system.
The former limit  corresponds to replacing the matrices $Y^\mu$ by commuting space-time coordinates $y^\mu$.   $y^0$ and $y^i,\;i=1,2$ can be regarded as time and space coordinates, respectively.  In addition, one replaces the commutator of functions of $Y^\mu$ by some Poisson bracket $\{\;,\;\}$  of the corresponding  functions of $y^\mu$. For this one introduces a noncommutativity parameter $\Theta$, and defines the commutative limit by $\Theta\rightarrow 0 $. 
To lowest order in $\Theta$,
$[{\cal F}(Y),{\cal G}(Y)]\rightarrow i\Theta \{{\cal F}(y),{\cal G}(y)\}$ for arbitrary functions ${\cal F}$ and ${\cal G}$.  Then (\ref{cmxpxmwx0}) goes to 
\be \{(y^1)^2+(y^2)^2,y^0\}= 0\label{pbypsymsyz}\;\ee in the limit.  This restricts to the spatial radius to being a  function of only the time $y^0$ coordinate
\be ( y^1)^2+ ( y^2)^2=a^2(y^0)\;,\label{cmtvcnstrnt}\ee
which defines a  rotationally invariant   manifold embedded in three space-time dimensions.
Similarly for $d=5$ we can write down an ansatz for  matrices which in the commutative limit yield four-dimensional rotationally
 invariant space-time manifolds.  They then are  possible candidates for a realistic cosmology.

Exact  solutions to matrix model equations of motion are notoriously difficult to obtain, which is true even for $d=3$.\cite{Arnlind:2012cx}  Exceptions are in  cases where the matrices define a finite dimensional Lie-algebra.  Well known examples of the latter are the fuzzy sphere
\cite{Madore:1991bw},\cite{Grosse:1994ed},\cite{CarowWatamura:1998jn},\cite{Alexanian:2000uz},\cite{Dolan:2001gn},\cite{Balachandran:2005ew},\cite{Iso:2001mg},\cite{Chaney:2015mfa} and  two-dimensional noncommutative de Sitter space.\cite{Ho:2000fy},\cite{Jurman:2013ota},\cite{Stern:2014aqa}  Here we show how one can generate a large class of rotationally invariant solutions to three-dimensional matrix models using a simple recursion relation.  While finding exact classical  solutions to matrix models can be nontrivial, it is easy to obtain solutions in the commutative limit.
  Starting with a  three and five space-time dimensional matrix models, we end up with two and four dimensional space-time manifolds, respectively,  in the commutative limit.  A large family of open, closed and  static space-time cosmologies can be recovered in this manner.   Among the matrix solutions is the Lorentzian fuzzy sphere discussed above which yields a closed two-dimensional universe in the commutative limit.  Other matrix solutions are obtained which resolve singularities  present in the commutative limit.  In the case of the five-dimensional matrix model, we were not able to obtain exact solutions.  However, a large family of solutions were obtained in the commutative limit.  Among them is four-dimensional de Sitter (or anti-de Sitter space) which is endowed with a nontrivial Poisson structure.

We examine the  bosonic sector of a three-dimensional
Lorentzian IKKT-type matrix model  in section two.  A totally antisymmetric cubic term is included in the action, which is consistent with the three-dimensional Lorentz symmetry. The dynamics is also invariant under  unitary gauge transformation   and translations. In the commutative limit of the theory it was possible to find a one parameter family of rotationally invariant solutions to the equations of motion.\cite{Stern:2014aqa} These solutions describe closed, open and static two-dimensional space-time surfaces with  some Poisson structure attached to the surface. For some special values of the parameter the Poisson brackets define  three-dimensional Lie algebras, and in these cases one can easily find exact solutions to the Lorentzian  matrix model.  They correspond to noncommutative de Sitter,  anti-de Sitter and static space-times, and have been discussed previously.\cite{Ho:2000fy},\cite{Chaichian:2000ia},\cite{Balachandran:2004yh},\cite{Jurman:2013ota},\cite{Stern:2014uea},\cite{Stern:2014aqa}    In section two we search for rotationally invariant  matrix solutions in  the generic case, i.e., solutions which are not in general  associated with any three-dimensional Lie algebra.  For this we must first give a definition of rotationally invariant matrices. After restricting to such matrix configurations we can obtain recursion relations for the eigenvalues of  matrices satisfying the equations of motion. The eigenvalues are discrete and define the spectra of the time and space (or radius) coordinates. The recursion relations are trivially solved  for noncommutative de Sitter and static space-times solutions.  More generally,  the recursion relations can  be solved numerically  and their  spectra describe   discrete versions of  open space-time universes. For the noncommutative de Sitter solution one has the principal, supplementary and discrete series representations of $su(1,1)$.\cite{Jurman:2013ota} One feature of the discrete series is that there is a minimum (maximum) time eigenvalue which is associated with the minimum radial eigenvalue. In the commutative limit it corresponds to an initial (final) space-time singularity. Thus the discrete series solution provides a noncommutative resolution of a big bang (crunch) singularity.

There are two disadvantages to the approach  described above for finding matrix solutions.  a) While the matrix analogues of  open space-time universes are easy to obtain, the solution for the matrix analogues of closed space-times universes
 is more problematic.  The recursion relations which are derived in section two require infinite dimensional matrix solutions, which may not be an appropriate assumption for modeling a closed noncommutative space-time.  In this regard, we are unable to find any finite dimensional  matrices which solve the 
Lorentzian IKKT-type   matrix model (containing only the additional cubic term in the action).  b) It was shown previously\cite{Stern:2014aqa}  that all the rotationally invariant solutions to that particular model have tachyonic-like excitations and so the stability of the solutions is not insured.  This is seen after taking the commutative limit.

The conclusions a) and b) change when additional terms are included in the matrix model action.  This is the case for a quadratic or mass-like term which we consider in section three.  The inclusion of this term preserves the unitary gauge symmetry and Lorentz symmetry, but breaks translation invariance. The modified equations of motion yield a multi-parameter family of rotationally invariant solutions  in the commutative limit.  They  can be solved for numerically, and  have novel features. Among them   are solutions which exhibit a transition from a rapid inflation to a non inflationary phase.  Another solution yields a closed universe   with an associated  $su(2)$ algebra (in contrast with the Lorentz symmetry of the embedding coordinates). Its matrix analogue is a Lorentzian  fuzzy sphere which has finite dimensional representations and describes a noncommutative closed universe.\cite{Chaney:2015mfa}

In section four we consider small perturbations about the rotationally invariant  matrix solutions and then take the commutative limit of the action.  The result is 
 a scalar field theory on the space-time manifold associated with the commutative solution.  The general analysis involves obtaining a nontrivial Seiberg-Witten map\cite{Seiberg:1999vs} from the noncommutative solution.  When the quadratic term is included in the matrix action we get that the effective mass-squared of the scalar field can be positive ensuring the stability of the  field theory in the commutative limit.

We examine a five-dimensional IKKT-type matrix model in section 5.  This system is physically relevant since its commutative limit can yield four dimensional space-times.  A totally antisymmetric fifth order term is included in the bosonic sector of the  matrix model action.  
The commutative limit has solutions describing four-dimensional  cosmologies.  One solution, which occurs in the limit that the Yang-Mills term vanishes, is four-dimensional de Sitter space $dS^4$.   The Poisson structure on these spaces preserves three dimensional rotation invariance. The   Seiberg-Witten map for the four-dimensional de Sitter solution  is applied to write the perturbative action in terms of commutative gauge fields and a scalar field.  Magnetic monopoles are shown to emerge from the perturbations. In section 5 we also define a notion of rotational symmetry for the five dimensional matrix model.   Finally, by changing the signature of the background metric we can obtain a four-sphere solution $S^4$ and   a four-dimensional anti-de Sitter solution  $AdS^4$ in the commutative limit of the five-dimensional matrix model. The  four-sphere solution is distinct from the commutative limit of the fuzzy four-sphere appearing in earlier works.\cite{Kimura:2002nq},\cite{Valtancoli:2002sm},\cite{Azuma:2004yg} The $AdS^4$ solution is also new.   For generic values of the parameters of the theory,  we find that the Poisson brackets  are nonvanishing at the  $AdS^4$ boundary.  An exception case is an $AdS^4$ solution which follow from an action which consists only of a totally antisymmetric term.   In that case the Poisson brackets vanish at the boundary, but not in the interior. 

Concluding remarks are made in section 6.

 In Appendix A we give the result for the Seiberg-Witten map on a general two-dimensional manifold. The Seiberg-Witten map on four-dimensional de Sitter space appear in Appendix B.

\section{Translational invariant Lorentzian IKKT-type model}

\setcounter{equation}{0}

Here we examine the bosonic sector of a Lorentzian IKKT-type matrix model in three space-time dimensions. The  dynamics for the  three infinite-dimensional Hermitean matrices $Y^\mu$, $\mu=0,1,2$ is determined from the action
\be S(Y)=\frac 1{g^2}{\rm Tr}\Bigl(-\frac 14 [Y_\mu, Y_\nu] [Y^\mu,Y^\nu]  {\color{black}-}\frac 23 i \alpha \epsilon_{\mu\nu\lambda}Y^\mu Y^\nu Y^\lambda\Bigr)\;,\label{mmactn}\ee  where a  totally antisymmetric  cubic term is added to  the standard Yang-Mills term and $\alpha$ and $g$ are constants. Our conventions are $\epsilon_{012}=1$, and we raise and lower indices with the flat metric $\eta_{\mu\nu}=$diag$(-1,1,1)$. The resulting equations of motion are
\be [ [Y_\mu,Y_\nu],Y^\nu] {\color{black}-}i\alpha \epsilon_{\mu\nu\lambda}[Y^\nu,Y^\lambda] =0\label{eqofmot} \ee
They are invariant under:

\noindent i)
Lorentz transformations $Y^\mu\rightarrow L^\mu_{\;\;\nu} Y^\nu$, where $L$ is a $3\times 3$ Lorentz matrix,

\noindent
ii) translations in the three-dimensional Minkowski space $Y^\mu\rightarrow Y^\mu+v^\mu\BI$, where $\BI$ is the unit matrix, and

\noindent iii) unitary `gauge' transformations, $Y^\mu\rightarrow UY^\mu U^\dagger$, where $U$ is an infinite dimensional unitary matrix.

Our interest shall be in constructing solutions to (\ref{eqofmot}) which are rotationally invariant in the 1-2 plane.  This will of course require defining a notion of rotationally invariant matrix configurations which we put off to subsection 2.2.1.  We first review the much simpler problem of finding rotationally invariant solutions in the commutative limit of the matrix model equations.

\subsection{Commutative limit}
The commutative limit of the matrix equations was examined previously in \cite{Stern:2014aqa} and a family of rotationally invariant solutions  were obtained. We review them here.  As stated in the introduction, the commutative limit corresponds to replacing the matrices $Y^\mu$ by commuting space-time coordinates $y^\mu$, and  the commutator of functions of $Y^\mu$ is replaced by some Poisson bracket $\{\;,\;\}$  of the corresponding  functions of $y^\mu$. 
The Poisson brackets on the three dimensional space spanned by $y^\mu$  are singular, and a function of the coordinates can be found which is central in the Poisson bracket algebra.  Setting that function equal to a constant  yields a two dimensional surface ${\cal M}_2$, upon which a nonsingular Poisson bracket can be defined. Similar arguments can be made to recover an even dimensional manifold starting with a $d=$odd dimensional matrix model.  Say that    $\tau$ and $\sigma$ parametrize the two-dimensional surface,  where   $\tau$ is a time-like parameter and $\sigma$ is space-like.   We will assume that any time slice of  ${\cal M}_2$ is a circle, $0\le\sigma<2\pi$.  In terms of the three embedding coordinates the surface is defined  by the functions $y^\mu=y^\mu(\tau, e^{i\sigma})$.  
Since  ${\cal M}_2$ is a two-dimensional surface the Jacobi identity is automatically satisfied, and  for any two functions ${\cal F}(\tau, e^{i\sigma})$ and ${\cal G}(\tau, e^{i\sigma})$ on ${\cal M}_2$ we can write
\be \{{\cal F},{\cal G} \}(\tau, e^{i\sigma})= h\,\Bigl( \partial_\sigma{\cal F} \partial_\tau {\cal G}-\partial_\tau {\cal F} \partial_\sigma {\cal G} \Bigr)\label{pbtauphi}\;,\ee where in general $h$ is some function of $\tau$ and $ e^{i\sigma}$. 

Since  the matrix model action (\ref{mmactn}) and the equations of motion (\ref{eqofmot}) can be expressed in terms of commutators, their commutative limit can be expressed in terms of Poisson brackets. In order
that all terms survive in the commutative limit, we need that $\alpha$ vanishes in the limit, more specifically, that it is proportional to $\Theta$. We write as $\, \alpha\rightarrow{\color{black}+}\upsilon \Theta$, with $ \upsilon\;\;{\rm finite}$. Then the commutative limit of the action is
\be
S_c(y)=\frac 1{g_c^2}\int_{{\cal M}_2} d\mu(\tau,\sigma)\Bigl(\frac 1{4} \{y_\mu, y_\nu\}\{y^\mu,y^\nu\} {\color{black}+}\frac {\upsilon }{3}\epsilon_{\mu\nu\lambda}\,y^\mu\{y^\nu, y^\lambda\}\Bigr)\;,\label{cmtvlmtsc}
\ee
where $g_c$ is the  commutative limit of the coupling $g$ and
$ d\mu(\tau,\sigma)$ is the integration measure on ${\cal M}_2$. The latter is required to be consistent with the cyclic trace identity,
\be\int_{{\cal M}_2} d\mu(\tau,\sigma)\,\{{\cal F},{\cal G}\}\,{\cal H}=\int_{{\cal M}_2} d\mu(\tau,\sigma)\,{\cal F}\,\{{\cal G},{\cal H}\}\;,\ee
 for arbitrary functions ${\cal F},\,{\cal G}$ and ${\cal H}$ on ${\cal M}_2$. We can then take \be d\mu(\tau,\sigma)=d\tau d\sigma/h\ee
The  commutative limit of the equations of motion (\ref{eqofmot}) is given by
\be \{\{y_\mu,y_\nu\},y^\nu\}{\color{black}-}\upsilon \epsilon_{\mu\nu\rho}\{y^\nu,y^\rho\} =0 \label{clmeas}
\,\ee
The dynamics retains its invariance under i)
Lorentz transformations,
ii) translations and
iii) gauge transformations.
Infinitesimal gauge variations have the form $\delta y^\mu=\Theta\{\Lambda,y^\mu\}$, where $\Theta$ again denotes the noncommutativity parameter and $\Lambda$ is an infinitesimal function on ${\cal M}_2$.

The dynamical equations coincide with string equations of motion. Here the relevant string action is
\be S_{string}=-{\cal T}\biggl[\int_{{\cal M}_2} d\tau d\sigma\, \sqrt{-{\tt g}}\;{\color{black}- }\;\frac{\upsilon}{3}\int _{{\cal M}_2} \epsilon_{\mu\nu\rho}y^\mu dy^\nu\wedge dy^\rho\biggr]\label{clnbns}\;,\ee where ${\tt g}$ is the determinant of the induced metric \be {\tt g}_{{\tt ab}}(\tau,\sigma)=\partial_{\tt a}y^\mu\partial_{\tt b} y_\mu\;,\qquad\;{\tt a}=\tau,\sigma\;\label{ndcdmtrc}\ee on ${\cal M}_2 $ and the constant ${\cal T}$ denotes the string tension. The first term in
(\ref{clnbns}) is the Nambu-Goto action, while the second corresponds to a coupling to a Neveu-Schwarz field of the from $B_{\mu\nu}\propto \epsilon_{\mu\nu\lambda}y^\lambda$. Both terms are reparametrization invariant, and respect Poincar\'e symmetry. They lead to the equations of motion
\be \Delta y_\mu{\color{black}- } 2\upsilon n_\mu=0\;\label{cleom}\ee Concerning the first term, $\Delta =-\frac 1{\sqrt{-{\tt g}}}\partial_{\tt a}\sqrt{-{\tt g}} {\tt g}^{\tt ab}\partial_{\tt b}$ is the Laplace-Beltrami operator on the world sheet, ${\tt g}^{\tt ab}$ denotes the components of the inverse induced metric, ${\tt g}^{\tt ab}{\tt g}_{\tt bc}=\delta ^{\tt a}_{\tt c}$. Concerning the second term, $n_\mu=\frac 1{2{\sqrt{-{\tt g}}}}\epsilon^{\tt ab}\epsilon_{\mu\nu\rho}\partial _{\tt a} y^\nu\partial_{\tt b} y^\rho$ is a space-like unit vector normal to the world sheet and $\epsilon^{\tau\sigma}=-\epsilon^{\sigma\tau}=1$. The string equations (\ref{cleom}) were shown to be identical to the equations (\ref{clmeas}) when the following condition is satisfied\cite{Arnlind:2012cx}
\be h =\frac 1{\sqrt{-{\tt g}}} \label{his1vrsmt}\ee

We denote solutions to the equations of motion by $y^\mu=x^\mu(\tau, e^{i\sigma})$, and focus on  solutions with an $SO(2)$ isometry group,  associated with rotations in the 1-2 plane.
For this we write the ansatz
\be \pmatrix{x^0\cr x^1\cr x^2}= \pmatrix{\tau\cr a( \tau)\cos\sigma\cr a( \tau)\sin\sigma}\;\label{ncprtztn}\ee
Here we have introduced a factor $a(\tau)$ which is the radius at any $\tau$-slice.  The ansatz (\ref {ncprtztn}) is consistent with (\ref{cmtvcnstrnt}).
The invariant interval on the surface is
\be ds^2=-(1-a'(\tau)^2)\,d\tau^2 +a(\tau)^2\,d\sigma^2\label{gnrl2dia}\;,\ee
the prime denoting differentiation in $\tau$.  This gives  the Ricci scalar
\be {\tt  R}=\frac{2 a''(\tau )}{a(\tau)\Big(1-a'(\tau)^2\Bigr)^2}\ee

Rotational  invariance in the 1-2 plane requires that we restrict $h$ in (\ref{pbtauphi}) to being a function of only $\tau$.
In order to have a solution to (\ref{clmeas}), the functions $a$ and $h$ need to satisfy
\be \Bigl ((aa'h)'+h-2\upsilon\Bigr)\,h=0\qquad \quad \Bigl(2ha'+ah'-2\upsilon a'\Bigr)\,ah=0\label{3difeqspct}\,\ee
 From these equations it follows that $h^2{\tt g}$ is a constant of integration, which is consistent with the condition (\ref{his1vrsmt}).
We can use (\ref{his1vrsmt}) to eliminate $h(\tau)$ and obtain a second order equation for the scale factor
\be \frac {a''}a= \Bigl(\frac{a'}a\Bigr)^2
-\frac 1{a^2}+\frac{ 2\upsilon} a (1-a'^2)^{\frac 32}\label{scndrdrfra}
\ee This yields the integral of the motion \be {\cal E}=a/{\sqrt{1-a'^2}}-\upsilon a^2\;,\label{calE}\ee which was shown in \cite{Stern:2014aqa} to be associated with the energy of a bosonic string. From (\ref{calE}) we then get the following Friedmann-type equation for $a(\tau)$:
\be \Bigl(\frac{a'}a\Bigr)^2-\frac 1{a^2}= -\frac 1{({\cal E}+\upsilon a^2)^2}\;\label{frdmneq}\ee 

Solutions to (\ref{frdmneq}) can be
expressed in terms of inverse elliptic integrals, which are plotted in figure 1.
\begin{figure}[placement h]
\begin{center}
\includegraphics[height=3.25in,width=3.0in,angle=0]{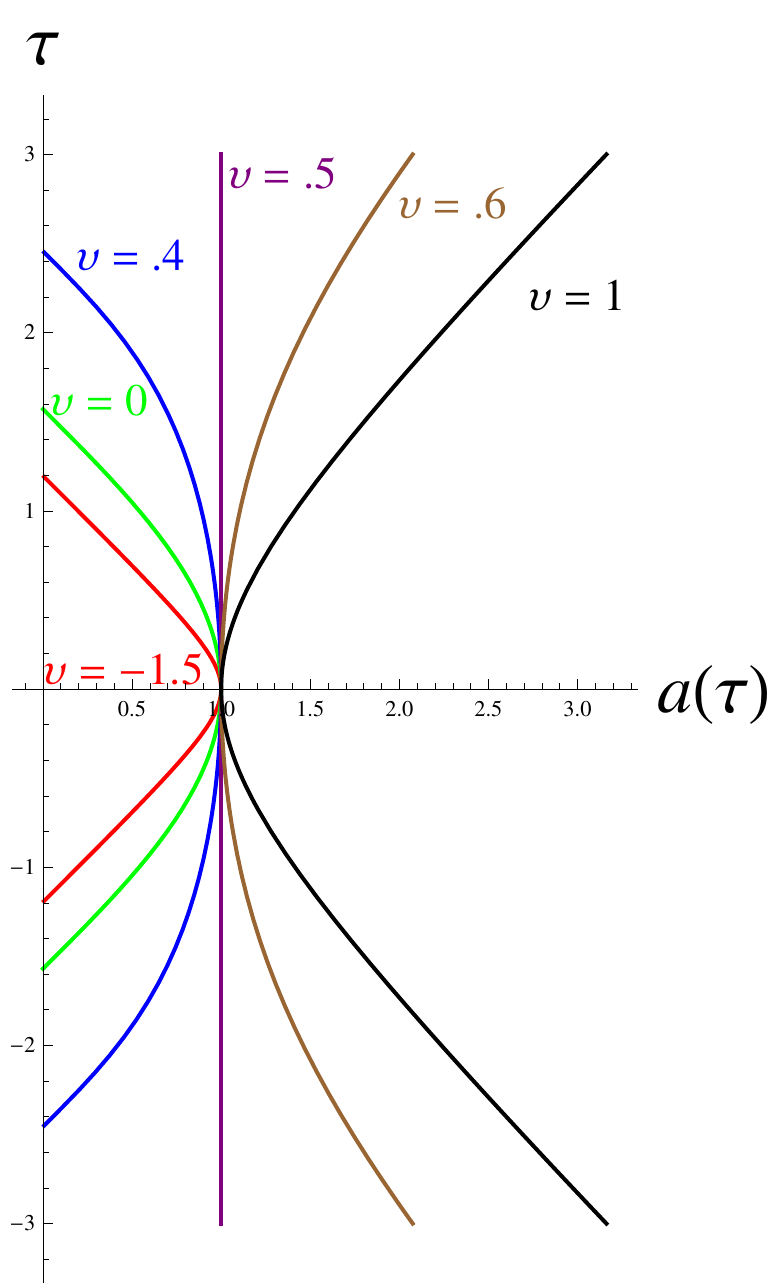}
\caption {Numerical solution to (\ref{scndrdrfra}) for $ \upsilon=-1.5,0,.4,.5,.6$ and $1$. $ \upsilon=.5$ and $1$ correspond to the cylinder and de Sitter solutions, respectively. The boundary values are $a(0)=1$ and $a'(0)=0$.
}
\end{center}
\end{figure}
For all the solutions plotted there (except the limiting case of $\upsilon=\frac 12$) we assume that $a$ has a turning point at $\tau=0$, i.e., $a(0)=1$ and $a'(0)=0$. The solutions describe closed, stationary and open space-times, the choice depending on  values for $\upsilon$.\cite{Stern:2014aqa} Closed two-dimensional space-times, having initial and final singularities, occur for $\upsilon<\frac 12$. The limiting case of $\upsilon=\frac 12$ gives the static or cylindrical space-time solution.\cite{Stern:2014uea} Open universe solutions correspond to $\upsilon>\frac 12$.

Exact expressions for the solutions exist for different values of $\upsilon$.  They are:

\smallskip
\noindent {\it a)}
For the case of $\upsilon= 0$, one has the simple expression
\be a(\tau)=\cos \tau\;,\qquad\quad- \frac \pi 2\le \tau\le\frac \pi 2 \;,\label{afhnqt}\ee 
where once again we assumed $a(0)=1$ and $a'(0)=0$, which  leads to singularities at $\tau=\pm\frac \pi 2$.
It defines the surface $ ( x^1)^2+ ( x^2)^2=\cos^2(x^0)$, with metric given by
\be ds^2=\cos^2 \tau\, (-d\tau^2+d\sigma^2) \;, \ee and Ricci curvature ${\tt R}=-2\sec^4 (\tau)$, the latter being singular at $\tau=\pm \frac\pi 2$.
The Poisson brackets of the embedding coordinates are
\be \{x^0,x^1\} =\frac { x^2}{\cos^2(x^0)}\qquad \{x^0,x^2\} =-\frac { x^1}{\cos^2(x^0)}\qquad \{x^1,x^2\}={\tan (x^0)}\;,\ee
where we used (\ref{pbtauphi}) and (\ref{his1vrsmt}).

\smallskip
\noindent {\it b)}
For $\upsilon=\frac 12$ the solution is simply
\be a=1\label{aeqcnst}\ee
The manifold is just a cylinder of unit radius with a flat metric tensor
\be ds^2=-d\tau^2+d\sigma^2 \ee
Using (\ref{pbtauphi}) and (\ref{his1vrsmt}) one now gets the Poisson brackets
\be \{x^0,x^1\} = x^2\qquad \quad \{x^0,x^2\} =- x^1\qquad \quad \{x^1,x^2\} =0\;,\label{cve2alg}\ee which define the three-dimensional Euclidean algebra.

\smallskip
\noindent {\it c)}
When $\upsilon= 1$ one gets
\be a(\tau)^2=1+\tau^2\;,\label{dScmtvesln}\ee
corresponding to a de Sitter space-time, \be ( x^1)^2+ ( x^2)^2-(x^0)^2=1\label{toodeeds}\ee
The invariant measure is given by
\be ds^2=-\frac {d\tau^2}{a(\tau)^2}+a(\tau)^2d\sigma^2 \;, \ee corresponding to yielding a constant positive Ricci curvature ${\tt R}=2$. The Poisson brackets on the surface define the $su(1,1)$ algebra
\be \{x^0,x^1\} = x^2\qquad \quad \{x^0,x^2\} =- x^1\qquad \quad \{x^1,x^2\} =-x^0\;\label{cmdS2alg}\ee

\noindent
Since the Poisson brackets for solutions {\it b)} and {\it c)} define Lie algebras, their noncommutative analogues are easy to obtain.  One simply replaces the Poisson brackets  by commutation relations. With the exception of these two cases, obtaining the matrix analogues of classical solutions is nontrivial. We give a procedure for finding `rotationally invariant' matrix solutions in the following subsections.

\subsection{Matrix solutions}
Here we search for matrix analogues of the  rotationally invariant solutions  of subsection 2.1 above to the commutative equations of motion (\ref{clmeas}). Our aim is to obtain the spectra of the matrices which solve the equations, which then give  lattice versions of the commutative solutions depicted in the plots in figure 1.  After first defining the meaning of rotational invariance for the matrices in subsection 2.2.1, we obtain recursion relations for the spectra in subsection 2.2.2.  Exact solutions to the recursion relations are discussed in subsections 2.2.3 and 2.2.4, and additional remarks concerning finite dimensional solutions and stability are made in subsections 2.2.5 and 2.2.6.

As a preliminary step it is convenient to write down an alternative expression for the  commutative solutions of  subsection 2.1. For this we utilize a different parametrization of the two dimensional manifolds. We replace $\tau$ by some  other time coordinate $t$, which along with $\sigma$, satisfies the fundamental Poisson bracket
\be \{\sigma,t\}=1\label{brbtwnsigat} \;,\ee
which has a simple noncommutative extension.
The previous commutative solutions can now be written as $y^\mu=x^\mu(t,e^{i\sigma})$.
We then regard $x^0$ and the scale factor, which we now denote by $\tilde a$, as functions of $t$, thereby replacing (\ref{ncprtztn}) with
\be \pmatrix{x^0\cr x^1\cr x^2}= \pmatrix{x^0(t)\cr \tilde a( t)\cos\sigma\cr \tilde a( t)\sin\sigma}\;\label{ncprtztnrpz}\ee
Then the equations of motion (\ref{clmeas}) give
\be 2\partial_t\tilde a\,(\partial_tx^0 {\color{black}-}\upsilon) +\tilde a \,\partial_t^2 x^0 =0\qquad\quad ( \partial_t x^0 {\color{black}-}\upsilon)^2-\upsilon^2+\partial_t(\tilde a \partial_t \tilde a)=0\ee
The first equation implies that
\be k=\tilde a^2\,(\partial_t x^0 {\color{black}-}\upsilon)\label{keqcnst}\ee is independent of $t$. The second equation then says that the dynamics $ \tilde a^2$ is determined by a simple force equation:
\be \frac 12 \partial_t^2(\tilde a^2)=-\frac {k^2}{\tilde a^4}+\upsilon^2\label{frceqasq}\ee
The solutions are characterized by $k$ and the conserved `energy' $\frac 14\Bigl(\partial_t (\tilde a^2)\Bigr)^2-\frac{ k^2}{\tilde a^2}-\upsilon^2\tilde a^2$. They are, of course, equivalent to those found previously in subsection 2.1. To see this one only needs to apply the reparametrization $t\rightarrow \tau=x^0(t)$.

\subsubsection{A rotationally invariant ansatz}

We now return to the matrix model described by the action (\ref{mmactn}). 
 Upon  defining $ Y_\pm$ as in (\ref{yplsy-}) the equations (\ref{eqofmot}) can be written according to
\beqa &&[Y_+,[Y_-,Y^0]]+\frac 12[Y^0,[Y_+,Y_-]]\;-\;\alpha [Y_+,Y_-] \;=\;0\cr&&\cr&&
[Y^0,[Y^0,Y_-]]+\frac 12[Y_-,[Y_-,Y_+]]+2\alpha [Y^0,Y_-]\;=\;0\label{eominAH}
\eeqa
We wish to write down a rotationally invariant ansatz for  the matrices $Y^0$ and $ Y_\pm$ which reduces to  (\ref{ncprtztnrpz})  in the commutative limit.   Different definitions are possible.
We require that our choice satisfies (\ref{cmxpxmwx0}). 
 Our  ansatz shall be expressed in terms of functions of two  infinite dimensional matrices $\hat t$ and  $ e^{i\hat \sigma}$, and are the matrix analogues of $ t$ and  $ e^{i\sigma}$, respectively.  The former is hermitean and the latter is unitary.  The matrix analogue of the Poisson bracket (\ref{brbtwnsigat}) is the  commutation relation
\be [ e^{i\hat \sigma},\hat t]=-\Delta e^{i\hat \sigma}\label{tthtacr}\;,\ee
where 
$\Delta$ is a central  element with  units of time  which is assumed to be linear in the noncommutative parameter. $ e^{i\hat \sigma}$ generates time translations $\hat t\rightarrow \hat t+\Delta$.   Together $ e^{i\hat \sigma}$ and $\hat t$ generate the algebra of the noncommutative cylinder.\cite{Chaichian:2000ia},\cite{Balachandran:2004yh},\cite{Stern:2014uea}

 For solutions to (\ref{eominAH}), which we denote by $Y^\mu=X^\mu$, we take 
\be X_+=X^1 + iX^2 =A(\hat t) e^{i\hat \sigma}\qquad\qquad X^0=X^0(\hat t)\label{WAeit}\,\ee
This is consistent with our definition (\ref{cmxpxmwx0}) of rotation invariance.
Here we restrict $X^0$ and $A$ to being real polynomial functions of $\hat t$.  Then $A(\hat t) $ and $X^0(\hat t)$ are infinite dimensional hermitean matrices.  In the commutative limit, the ansatz (\ref{WAeit}) agrees with the  expression (\ref{ncprtztnrpz}).   After substituting the ansatz  into (\ref{eominAH}) one gets
\beqa \Bigl( X^0(\hat t)-X^0(\hat t-\Delta)-\alpha\Bigr)A(\hat t)^2
- \Bigl( X^0(\hat t+\Delta)-X^0(\hat t)-\alpha\Bigr)A(\hat t+\Delta)^2 =0
&&\cr&&\cr
\frac 12 \Bigl( A(\hat t-\Delta)^2+A(\hat t+\Delta)^2-2A(\hat t)^2 \Bigr)+\Bigl( X^0(\hat t)-X^0(\hat t-\Delta)-\alpha\Bigr)^2-\alpha^2=0&&
\label{useBCH}
\eeqa
The first equation states that $\Bigl( X^0(\hat t)-X^0(\hat t-\Delta)-\alpha\Bigr)A(\hat t)^2$ is invariant under discrete translations $\hat t\rightarrow \hat t+ n\Delta$, $n=$integer, and is the matrix analogue of (\ref{keqcnst}).

\subsubsection{Recursion relations}
We next write down recursion relations for the eigenvalues of the matrices $ X^0(\hat t)$ and $A(\hat t)$.
The spectrum for the operator $\hat t$ is discrete, with equally spaced eigenvalues
\be t_n= t_0- n\Delta
\;,\qquad n\in
{{Z}}\label{dsctsptmt}\;,\label{tmspctrm}\ee where $t_0$ is real.
This follows since from the commutation relations (\ref{tthtacr}), $e^{2\pi i\hat
t/\Delta}$ is a central element. It is a constant phase $e^{2\pi it_0/\Delta}\BI$ in any
irreducible representation of the algebra, from which (\ref{tmspctrm}) results.

 The  eigenvalues of $ X^0(\hat t)$ and $A(\hat t)$ are real and we denote them by
\be x^0_n= X^0(t_n)\qquad\qquad a_n= A(t_n)\;, \label{x0nan}\ee
From (\ref{WAeit}), $X_+$ and $X_-$ act as lowering and raising operators, respectively, on the corresponding eigenvectors.
Since the eigenvalues of $X_1^2+X_2^2=\frac 12(X_+X_-+X_-X_+)$ are positive definite, we get that $ a_n^2+a_{n-1}^2\ge 0 $, for all $n$.
However, since we want $A(\hat t) $ to be hermitean, we get the stronger condition that
\be a_n^2\ge 0\;,\label{neqltyepj}\ee for all $n$.
From the equations of motion (\ref{useBCH}) we get the following recursion relations for the eigenvalues:
\beqa \Bigl( x^0_n- x^0_{n+1}-\alpha\Bigr)a_n^2
- \Bigl(x^0_{n-1}-x^0_n-\alpha\Bigr)a_{n-1}^2 =0
&&\cr&&\cr
\frac 12 \Bigl(a_{n+1}^2+a_{n-1}^2-2a_n^2 \Bigr)+\Bigl(x^0_n-x^0_{n+1}-\alpha\Bigr)^2-\alpha^2=0&&\label{rcrsnrltnah}
\eeqa
From the first equation, $k= \Bigl( x^0_n- x^0_{n+1}-\alpha\Bigr)a_n^2$ is independent of $n$, and then from the second equation we get a recursion relation for just $a_n$
\be \frac 12 \Bigl(a_{n+1}^2+a_{n-1}^2-2a_n^2 \Bigr)+\frac {k^2}{a_n^4}-\alpha^2=0\;,
\label{rcrsnrlfrantnah}
\ee
which is valid provided $a_n$ doesn't vanish. (\ref{rcrsnrlfrantnah}) is the lattice version of (\ref{frceqasq}).  Given the values for any neighboring pair of eigenvalues for $A$, we can determine the the entire series $\{a_n\}$. Then starting with one time eigenvalue, we can determine all of $\{x^0_n\}$ using  $ x^0_n- x^0_{n+1}=\alpha+ k/a_n^2$ .

Solutions are plotted in figure 2 for $ \alpha=.5,.51$ and $.6$ with $k=.5$ and boundary values $a_0=a_1=1$, $x^0_0=0$. $ \alpha=.5$ corresponds to the noncommutative cylinder solution which we discuss in subsection 2.2.3. $ \alpha=.51$ and $.6$ are examples of discrete versions of open universe solutions. Another example of a  discrete  open universe is the noncommutative de Sitter solution which corresponds to $k=0$. We discuss this case in subsection 2.2.4.
\begin{figure}[placement h]
\begin{center}
\includegraphics[height=2.75in,width=3in,angle=0]{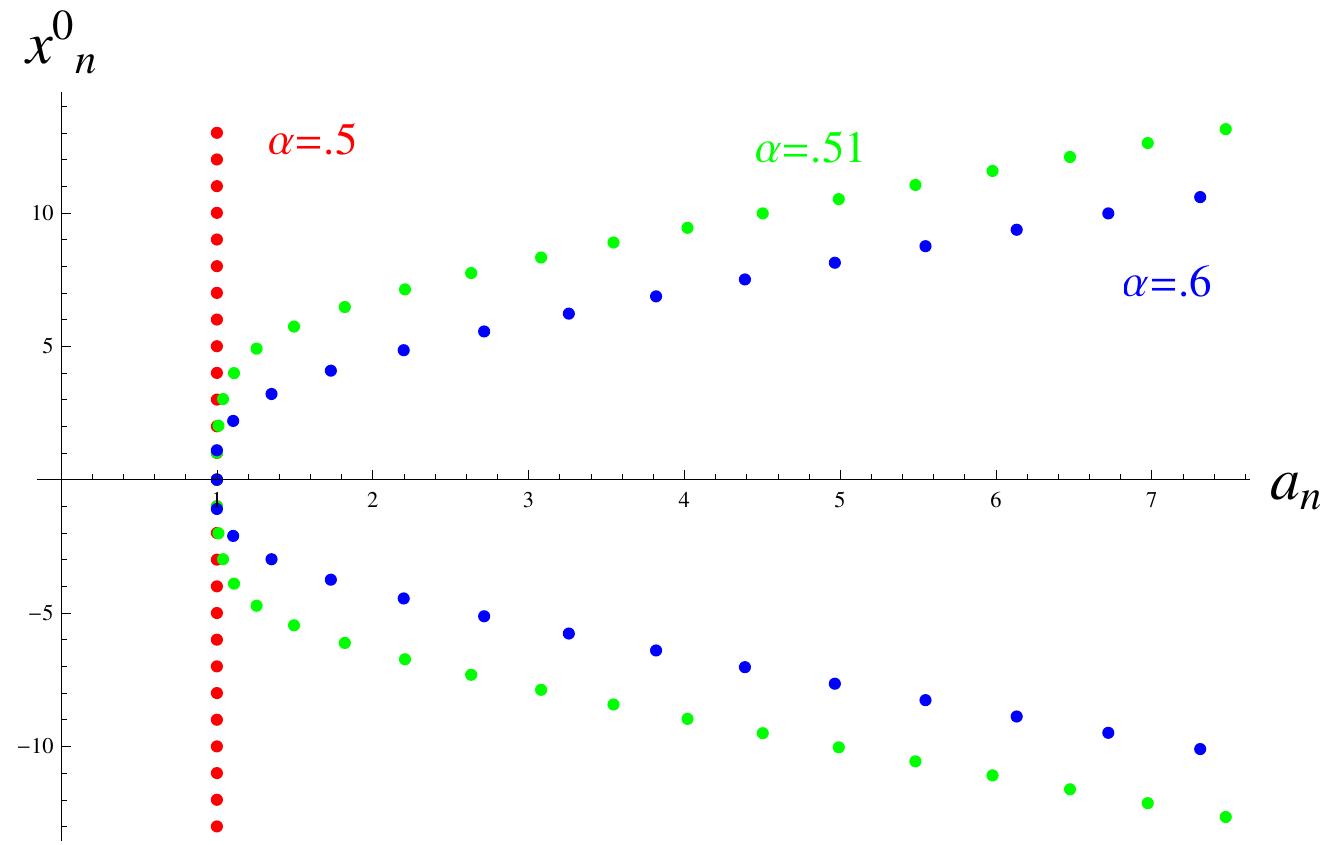}
\caption {Solutions to the recursion relation (\ref{rcrsnrlfrantnah}) for $ \alpha=.5,.51$ and $.6$ with $k= \Bigl( x^0_n- x^0_{n+1}-\alpha\Bigr)a_n^2$ fixed to be $.5$. $ \alpha=.5$ corresponds to the noncommutative cylinder solution, while $ \alpha=.51$ and $.6$ are examples of noncommutative analogues of open universe solutions. The initial conditions are $a_0=a_1=1$ and $x_0^0=0$.
}
\end{center}
\end{figure}
While figure 2 shows matrix analogues of the cylindrical and open space-time solutions, here we are unable to obtain matrix analogues of closed space-times. Related to this issue is the absence of solutions having $ \alpha<.5$ (or more generally, $ |\alpha |<|k|,$ along with initial conditions $a_0=a_1=1$). For these cases,
$a^2_n$ decreases to zero as one goes away from the initial values. ($a^2_n=0$ is analogous to zero radius in the continuous case; i.e.,  a cosmological singularity.)  As $a^2_n$ decreases to zero, the $ {k^2}/{a_n^4}$ term dominates in the recursion relation (\ref{rcrsnrlfrantnah}), i.e., the leading term in the expression for $a^2_{n+1}$ goes like $- {k^2}/{a_n^4}<0$. Then for some $n$, $a^2_n$ becomes negative, which is inconsistent with hermiticity. Thus, either such solutions do not exist or there must be raising or lowering operators that kill all states with $a^2_n<0$.
An example of the latter is  the discrete series representation of the de Sitter solutions, which is discussed in subsection 2.2.4.

We note that the analysis leading to recursion relations (\ref{rcrsnrltnah}) and (\ref{rcrsnrlfrantnah})
 is only valid for infinite dimensional solutions to the matrix equations, and moreover when the index  $n$ spans all positive and negative integers. (\ref{WAeit}) is not valid if this is not the case. Alternatively, if $n$ does not span all positive and negative integers it still may be possible to write
\be X_+=AU\;,\label{WAU}\ee
where $A$ and $U$ are diagonal and unitary matrices, respectively, and $X^0$ is a diagonal matrix. This is a  generalization of the ansatz (\ref{WAeit}). Both (\ref{WAeit}) and (\ref{WAU}) imply that $X_+X_-$ commutes with $X^0$, and so they  are consistent with the definition (\ref{cmxpxmwx0}) of rotational invariance.  Then $X_+X_-$ and $X^0$ have common eigenvalues.     In the case where  (\ref{WAeit}) holds they are, respectively, $a_n^2$ and $x^0_n$.
Even if (\ref{WAeit}) does not hold, it may still be possible that the recursion relations (\ref{rcrsnrltnah}) for the eigenvalues $x^0_n$ and $a_n^2$ of $X^0$ and $A^2$, respectively, are valid after restricting the values of the label $n$ in some fashion. For example, we find this  to be the case  for the discrete series of the de Sitter solutions, as is discussed  in subsection 2.2.4.

We next review well known examples of rotationally invariant matrix model solutions, which are exact solutions of the recursion relations (\ref{rcrsnrlfrantnah}).

\subsubsection{Noncommutative cylinder\cite{Chaichian:2000ia},\cite{Balachandran:2004yh},\cite{Stern:2014uea}}
A trivial solution of the recursion relation (\ref{rcrsnrlfrantnah}) is
\be x^0_{n}= -2\alpha n + x^0_0 \qquad\qquad a_n=a_0\qquad \qquad\label{nccylltc}\ee
where $k=\alpha a_0^2$ and $n\in Z$. $x^0_0$ and $a_0$ are real and here are identified with eigenvalues of $X^0$ and $A$ for the noncommutative cylinder. The solution represents the discrete version of the constant solution for $a$ (\ref{aeqcnst}). The noncommutative cylinder solution $Y^\mu=X^\mu$ is defined by the commutation relations
\be [X^0,X_+]=2\alpha X_+ \qquad\qquad [X_+,X_-]=0\;,
\ee
from which one recovers Poisson brackets (\ref{cve2alg}) in the  limit $\alpha\rightarrow 0$.
$x^0_{n}$ and $ a_n$ in (\ref {nccylltc}) are the eigenvalues, respectively, of $X^0$ and the square root of
$X_+X_-$, which is central in the algebra. The latter is constant in any irreducible representation of the algebra and is the radius-squared of the noncommutative cylinder. $X_+$ and $X_-$ are raising and lowering operators, respectively, for the eigenvectors of $X^0$.

\subsubsection{Noncommutative $dS^2$\cite{Jurman:2013ota}}
Another solution of the recursion relations is
\be x^0_{n}= -\alpha( n + \epsilon_0) \qquad\qquad a_{n}^2=\alpha^2n( n +2\epsilon_0+1 )+ a_0^2\label{ncds2ltc}\ee
where $\epsilon_0$ and $a_0$ are real. Now $k=0$.
This solution can be identified with  noncommutative (or fuzzy) $dS^2$ and it corresponds to the matrix analogue of the solution (\ref{dScmtvesln}). The relevant commutation relations for the matrix solution $Y^\mu=X^\mu$  now define the  $su(1,1)$ Lie algebra
\be [X^0,X_+]=\alpha X_+ \qquad\qquad [X_+,X_-]=-2\alpha X^0 \;,\label{suoneone}
\ee and they yield the Poisson brackets  (\ref{cmdS2alg}) in the $\alpha\rightarrow 0$ limit.

Irreducible representations of the $su(1,1)$ Lie algebra are well known and classified by eigenvalues of the central operator
\be \;R^2=\frac 12(X_+X_-+X_-X_+)-(X^0)^2 \; \label{copsu11}\;,\ee which we denote by $-\alpha^2j(j+1)$, along with $\epsilon_0$. $R$ is the length scale of the noncommutative de Sitter space. States $|j,\epsilon_0,n>$ in any irreducible representation can be taken to eigenvectors of $X^0$, with $X_+$ and $X_-$ behaving as lowering and raising operators, respectively,
\beqa X^0|j,\epsilon_0,n>&=&-\alpha (\epsilon_0+n)|j,\epsilon_0,n>\cr &&\cr
X_+|j,\epsilon_0,n>&=&i\alpha (j+\epsilon_0+n)|j,\epsilon_0,n-1>\cr &&\cr
X_-|j,\epsilon_0,n>&=&i\alpha (j-\epsilon_0-n)|j,\epsilon_0,n+1>\label{HWWdonsts}\eeqa
It follows that \be X_+X_-|j,\epsilon_0,n>=\alpha^2\Bigl( (\epsilon_0+n)(\epsilon_0+n+1)-j(j+1)\Bigr)|j,\epsilon_0,n>\;,\label{XpXmsu11} \ee
If we assume (\ref{WAeit}), or more generally  (\ref{WAU}),  then we can identify  $X_+X_-$ with $A^2$, with eigenvalues $a_n^2\ge 0$. Then comparing  (\ref{ncds2ltc}) with (\ref{XpXmsu11}) gives
\be a^2_0=\alpha^2\Bigl(\epsilon_0(\epsilon_0+1)-j(j+1)\Bigr)\ee
The inequality (\ref{neqltyepj}) in this case leads to
\be \Bigl(\epsilon_0+n+\frac 12\Bigr)^2\ge \Bigl(j+\frac 12\Bigr)^2 \label{e0n2jjp1}\;,\ee for all $n$.

Nontrivial representations are known to  fall into three categories: principal, supplementary and discrete series. For the principal and supplementary series, neither $j+\epsilon_0$ nor $j-\epsilon_0$ are integers, so that no states $|j,\epsilon_0,n>$ are killed by either $X_+$ or $X_-$. There are then no restrictions on the integers $n$ labeling the states.  One takes $j=-\frac 12 +i\rho$, with $\rho$ real, for the principal series, which identically satisfies (\ref{e0n2jjp1}). $j$ is assumed to be real for the supplementary series. Then if we choose $-\frac \pi 2\le \epsilon_0<\frac \pi 2$, we need that $|j+\frac 12|\le |\epsilon_0+\frac 12|$.

Finally, for the discrete series one has that either $j+\epsilon_0$ or $j-\epsilon_0$ are integers. For the former, we can choose $j+\epsilon_0=0$. Then from (\ref{HWWdonsts}), $X_+$ kills $|j,-j,0>$, which then serves the role as the bottom state for the irreducible representation $D^+(j) $.  In this case  $n$ is restricted to positive integers, including $0$. The inequality (\ref{e0n2jjp1}) is satisfied for $j\le0$. The resulting spectra for $X^0$ and $X_+X_-$ is given by
\be x^0_n=\alpha(j-n)\qquad \qquad a_n^2=\alpha^2(n+1)(n-2j)\;, \qquad n=0,1,2,... \;\label{sptrHWWd}\ee The time takes on only negative eigenvalues, assuming $\alpha>0$.
Similarly, if one chooses $j-\epsilon_0=0$, then from (\ref{HWWdonsts}), $X_-$ kills $|j,j,0>$. The latter serves the role as the top state for the irreducible representation $D^-(j)$ and   in this case  $n$ is restricted to negative integers, including $0$.    The inequality (\ref{e0n2jjp1}) is again satisfied for $j\le0$. The spectrum for $X_+X_-$ is the same as in the previous case (\ref{sptrHWWd}), while there is a sign flip for $x^0_n$, i.e., the signs are now all positive. We note that
(\ref{WAeit}) is not valid for the discrete solutions since the $n$ does not span all integers.
[The recursion relations (\ref{rcrsnrltnah}) are still valid, however, provided that we now restrict the integer $n$ in these relations to $n\ge 1$ for $D^+(j) $, and $n\le- 1$ for $D^-(j) $.] Furthermore, its generalization (\ref{WAU}) does not hold in general either. Only for $j=-\frac 12$ can we write $X_+$ in the form (\ref{WAU}). For $D^+(-\frac 12)$ the matrices $A$ and $U$ are given simply by
\be A_{nm}= \alpha (n+1)\delta_{n,m}\qquad\quad U_{nm}=i\delta_{n+1,m}\ee 

  We remark that the existence of a bottom (top) state for the discrete series, where there is a corresponding minimum value for the radial eigenvalues $a_n$, means that there is an initial (final) state.  It is thus a discrete analogue of a cosmological singularity.

Plots of the eigenvalues $x^0_n$ and $a_n$ of the matrices $X^0$ and $A$, respectively, are given in figure 3 for the discrete series $D^-(-\frac 12)$, as well as for the principal series with $\rho =10$ and $20$. 
\begin{figure}[placement h]
\begin{center}
\includegraphics[height=3.25in,width=3in,angle=0]{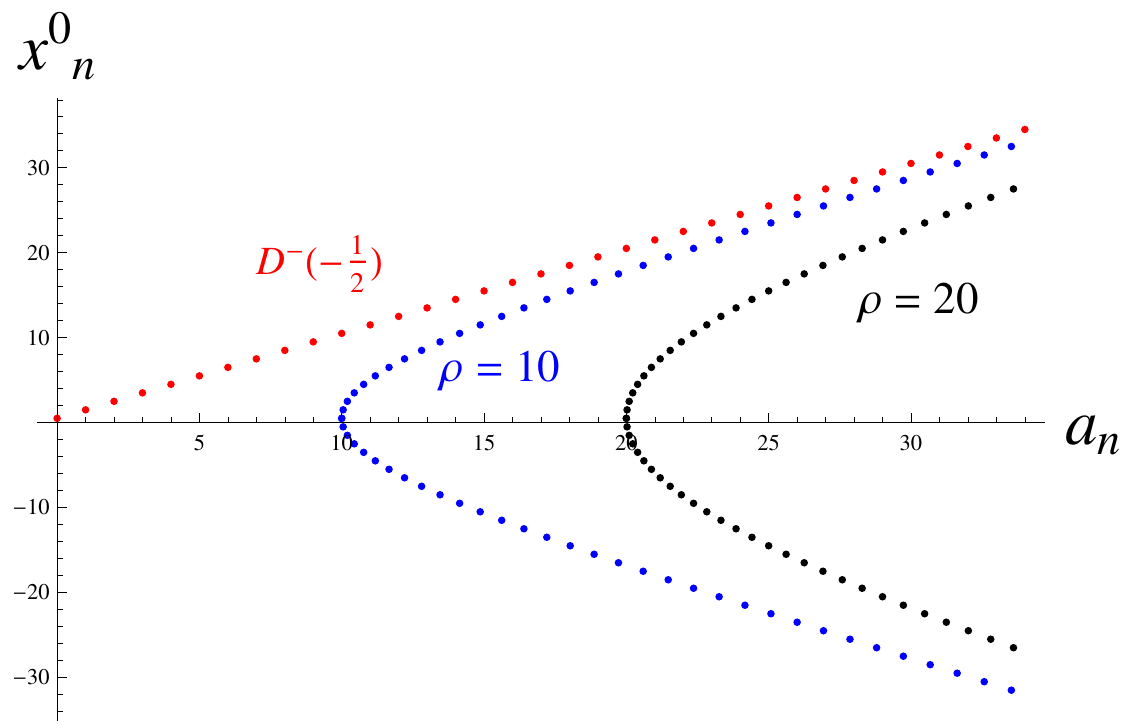}

\caption {Noncommutative de Sitter solutions. Plots of the eigenvalues $x^0_n$ and $a_n$ for the discrete series $D^-(-\frac 12)$ and for the principal series with $\rho =10$ and $20$. $\alpha=1$ and $\epsilon_0=-\frac 12$ is chosen in all cases.}
\end{center}
\end{figure}

\subsubsection{The question of finite dimensional solutions}

In the above example, it is well known that there are no finite dimensional  solutions of noncommutative  de Sitter space since there are no finite dimensional unitary representations of the $SU(1,1)$ group. More generally, one can ask whether or not there exist nontrivial finite dimensional matrix solutions of the equations of motion (\ref{eqofmot}). This question is relevant for knowing whether or not there are matrix solution analogues of the closed space-time cosmologies.  The latter are expected to emerge upon taking the $N\rightarrow \infty$ limit of the $N\times N$ matrix solutions, along with initial and final singularities on the resulting space-time manifold.  So for example, one can ask if there is a matrix analogue of the closed space-time solution (\ref{afhnqt}) of the commutative equations of motion (\ref{clmeas}).

As stated above, if $a_n^2$ tends to zero, it becomes necessary to terminate the series generated by the recursion relations (\ref{rcrsnrlfrantnah}) in order to prevent $a_n^2$ from becoming negative.  There must then exist a bottom or top state, which would correspond, respectively, to an initial or final singularity in the continuum limit.  Any matrix analogue of a closed space-time solution must have {\it both} a bottom and top state, and thus the matrix solution should  be  finite dimensional.
In this regard, we have not been able to find any nontrivial finite dimensional matrix solutions to (\ref{eqofmot}), and thus here we do not have matrix model analogues of the closed space-time solutions of subsection 2.1; i.e., all the solutions of (\ref{scndrdrfra}) with $\upsilon<\frac 12$.

Although we do not have a  proof that there are no nontrivial $N\times N$  solutions, for arbitrary finite $N$,  to the matrix equations (\ref{eqofmot}), it is easy to show that no nontrivial solutions exist for  the simplest case of $N=2$.
In that case we can set $X^\mu$ equal to a linear combination of Pauli matrices, one of which, say $X^0$, we can take up to a factor to be $\sigma_3$.
(Terms in $X^\mu$ which are proportional to the identity matrix  trivially solve the equations of motion.) $$
X^0=\sigma_3\qquad X^1=u_i\sigma_i\qquad X^2=v_i\sigma_i\;, $$
where $u_i$ and $v_i$ are real.  Here we are not making any additional restrictions such as rotational invariance.
Upon substituting into the equations of motion (\ref{eqofmot}) one gets
$$ \; u_3 u_i +v_3 v_i-\Bigl(\vec u^2 + \vec v^2 \Bigr)\delta_{i3}-\alpha \epsilon_{ijk}u_jv_k =0 \quad $$
$$ \vec u\cdot \vec v\; u_i- v_3\delta_{i3} -\Bigl(\vec u^2-1\Bigl) v_i-\alpha \epsilon_{ij3}u_j =0 $$
$$ \vec u\cdot \vec v\; v_i- u_3\delta_{i3} -\Bigr(\vec v^2-1\Bigl) u_i+\alpha \epsilon_{ij3}v_j =0 $$ The only real solutions are $u_i= u_3\delta_{i3},\;v_i= v_3\delta_{i3}$, but these are trivial solutions since then all $X^\mu$ are proportional to $\sigma_3$. Thus there are no nontrivial $2\times 2$ matrix solutions of the equations of motion
(\ref{eqofmot}).

\subsubsection{The question of stability}

Issues related to the stability of the rotationally invariant  solutions to (\ref{eqofmot}) were examined in \cite{Stern:2014aqa}. More specifically, \cite{Stern:2014aqa} was concerned with small perturbations about the  solutions  to (\ref{frdmneq}).   Leading order effects were examined upon perturbing in the noncommutative parameter  $\Theta$, or equivalently  $\alpha$. 
The perturbations about the solutions were expressed in terms of an abelian gauge field and scalar field (or nonabelian gauge fields and $N$ scalar fields if one expands about a stack of $N$ coinciding branes). This could be done in general with the use of an appropriate Seiberg-Witten map\cite{Seiberg:1999vs} on the noncommutative space-time associated with the solution.
Gauge transformations correspond to area preserving coordinate transformations on the two-dimensional surface, while the scalar field is associated with perturbations normal to the surface.  At leading order in  $\Theta$, the resulting perturbed action yielded the usual dynamics for a scalar field coupled to the gauge field on the two-dimensional commutative manifold.  Since  gauge fields are  nondynamical in two-space-time dimensions they can be eliminated leaving only the scalar field degree of freedom.
For all  values of the parameter $\upsilon$ appearing in the commutative theory, the remaining scalar field was found to be tachyonic.

The persistence of tachyonic modes and  the absence of any finite-dimensional solutions appears to be  generic features of the Lorentzian IKKT-type matrix model whose dynamics follows from the action (\ref{mmactn}).
On the other hand, they no longer are the case
when additional terms are included in the action.   For example it was recently found in \cite{Chaney:2015mfa}  that finite dimensional matrix solutions  exist when a  quadratic, or mass, term is added to the matrix model
 action.   With the same quadratic term the scalar field resulting from perturbations about the rotationally invariant solutions can have a positive mass-squared, thus ensuring stability of the commutative field theory.   We more generally explore the consequences of including the quadratic term  in  the following section.

\section{Inclusion of a quadratic term}

\setcounter{equation}{0}
We now  add a quadratic, or mass, term to (\ref{mmactn}).  The total matrix model action is then
\beqa S_{\tiny {\rm total}}(Y)&=&S(Y)+\frac \beta{2 g^2}{\rm Tr}Y_\mu Y^\mu\cr&&\cr &=&\frac 1{g^2}{\rm Tr}\Bigl(-\frac 14 [Y_\mu, Y_\nu] [Y^\mu,Y^\nu] {\color{black}-}\frac 23 i \tilde\alpha \epsilon_{\mu\nu\lambda}Y^\mu Y^\nu Y^\lambda+\frac \beta 2Y_\mu Y^\mu\Bigr)\;,\label{mmactnplsqd}\eeqa
where $\beta$ is a real constant and we now denote the coefficient of the cubic term by $\tilde\alpha$ in order to distinguish it from the noncommutative parameter $\alpha$ appearing in the rotationally invariant 
classical solutions of section 2.2.
The matrix equations of motion now read
\be [ [Y_\mu,Y_\nu],Y^\nu] {\color{black}-}i\tilde\alpha \epsilon_{\mu\nu\lambda}[Y^\nu,Y^\lambda] =-\beta Y_\mu\label{eomwthqdtrm}\ee
The i) $2+1$ Lorentz symmetry of the background space, as well as iii) the unitary gauge symmetry, is preserved by the last term, but ii) translation symmetry is broken when $\beta\ne 0$.

This system contains new solutions, as well as some of the previous solutions (even when $\beta\ne 0$).   Before discussing the matrix solutions, we once again find it convenient to first examine solutions in the commutative limit of the matrix model.

\subsection{Commutative limit}

We first write down the modification of the commutative  equations of subsection 2.1.  Now a much larger family of solutions exist.  We shall  give a (mostly) qualitative discussion of these solutions.
 
The commutative limit of the matrix model action can once again be expressed using the Poisson bracket (\ref{pbtauphi}) on some two-dimensional manifold ${\cal M}_2$.  In order for the cubic term in the action to survive in the limit we again need for its coefficient to be linear in the noncommutativity parameter $\Theta$, i.e., $ \tilde\alpha\rightarrow{\color{black}+} \upsilon \Theta$.
The quadratic term in the action will survive in the limit provided that $\beta$ goes like $\Theta^2$, i.e.,  $ \beta\rightarrow \omega \Theta^2$, with $ \omega\;\;{\rm finite}$. Then (\ref{cmtvlmtsc}) is replaced by
\be
S_c(y)=\frac 1{g_c^2}\int_{{\cal M}_2} d\mu(\tau,\sigma)\Bigl(\frac 1{4} \{y_\mu, y_\nu\}\{y^\mu,y^\nu\} {\color{black}+ }\frac {\upsilon }{3}\epsilon_{\mu\nu\lambda}\,y^\mu\{y^\nu, y^\lambda\}+\frac \omega 2 y_\mu y^\mu\Bigr)\;,\label{cmtvlmtscbne0}
\ee
where
$ d\mu(\tau,\sigma)=d\tau d\sigma/h$ is once again the invariant integration measure on ${\cal M}_2$. Not surprisingly, translational invariance is broken when $\omega\ne 0$.
The resulting equations of motion are now
\be \{\{y_\mu,y_\nu\},y^\nu\}{\color{black}- }\upsilon \epsilon_{\mu\nu\rho}\{y^\nu,y^\rho\} =\omega y_\mu \label{clmeaswne0}
\;\ee

Upon substituting the rotationally invariant expression (\ref{ncprtztn}) into (\ref{clmeaswne0}), we get the following equations for
$a$ and $h$, both of which are assumed to be functions of only $\tau$
\be (aa'h)'h+h^2-2\upsilon h=\omega\qquad \quad 2h(h-\upsilon) a'a+a^2h'h=\omega\tau\label{3difeqspctwne0}\;,\ee
the prime again denoting a derivative in $\tau$.
It follows that $h^2{\tt g}- \omega(a^2-\tau^2)$ is a constant of integration, where once again ${\tt g}$ is the determinant of the induced metric. We then get an explicit formula for $h$, and hence the integration measure
\be h=\sqrt{\frac{c_1- \omega(a^2-\tau^2)}{(1-a'^2)a^2}}\;,\label{dynfnfrh}\ee
$c_1$ being the constant of integration. Here we see that   the measure is not simply expressed in terms of metric, except for the case $\omega=0$ where we recover the result (\ref{his1vrsmt}) of the previous section.  For all $\omega$ and $\upsilon$, $h$ can be eliminated from the differential equation for $a$, which can be written
\be a''-\frac{a'^2}a+\frac 1a-\frac{2\upsilon}{ah}(1-a'^2)-\frac\omega{h^2}\Bigl(\frac \tau a\Bigr)'=0\label{fivefive}\;,\ee
generalizing (\ref{scndrdrfra}).   The breaking of time translation symmetry when $\omega\ne 0$ implies the absence of a conserved energy, and consistent with that, we have not found a generalization of the quantity  (\ref{calE}) which is conserved when $\omega=0$.

There is now a large family of solutions, including those discussed in 2.1 when $\omega= 0$.  
Among them are some exact solutions, all of which have $h$ equal to a constant value: 

\begin{enumerate}

\item There are two distinct $dS^2$ solutions to (\ref{3difeqspctwne0}) of the form (\ref{dScmtvesln})  when  $\upsilon^2+2\omega >0$ and $\omega\ne 0$.  [Here we assume  $\upsilon$ and  $\omega$ are finite.] They yield the following constant values for $h$,
\be h_\pm=\frac 12(\upsilon\pm \sqrt{\upsilon^2+2\omega})\;\label{2dstnctds2}\ee
The  solution is degenerate when $\upsilon^2+2\omega=0$, and no de Sitter solution exists for $\upsilon^2+2\omega<0$. 

\item
A $dS^2$ solution exists to the equations of motion   in the limit  $\upsilon$, $\omega\rightarrow\infty$, with $\frac \omega\upsilon$ finite and nonzero.   In this limit, the  kinetic energy (or Yang-Mills) term is  absent from  the action (\ref{cmtvlmtscbne0}).  In this case
\be h=-\frac{\omega}{2\upsilon}\label{2ddsnke}\ee
If both the kinetic energy term and the quadratic term are absent from the action  and only the totally antisymmetric term remains, i.e.,  $\upsilon\rightarrow\infty$ and  $\frac \omega\upsilon\rightarrow 0$, then there are only trivial solutions to the equations of motion,  $\{x^\mu,x^\nu\}=0$.  In the matrix model, all  matrices $X^\mu$ commute in this case.  This result does not generalize to higher dimensions where one can have nontrivial solutions of the equations of motion when only a totally antisymmetric term appears in the action, as we show in subsection 5.4.2.

\item
Another solution, which exists only  when $\omega\ne0$, is a sphere, $ ( x^1)^2+ ( x^2)^2+(x^0)^2=1$, embedded in three-dimensional Minkowski space-time.\cite{Chaney:2015mfa}  The solution is
\be a(\tau)^2=1-\tau^2   \qquad\qquad h=2\upsilon\;,\qquad\qquad -1\le \tau\le 1\;,\label{sphLntz}\ee  which is only valid for $\omega=-4 \upsilon^2$.
(More generally, by introducing another real parameter  it is deformed to an ellipsoid   embedded in three-dimensional Minkowski space-time.)   The invariant measure obtained from the induced metric is
\be ds^2=-\frac{1- 2\tau^2}{1- \tau^2}  d\tau^2+(1- \tau^2)d\sigma^2 \;,\ee  which
differs from that of the Euclidean sphere.  The metric tensor does not have  definite signature;  It is  Lorentzian for  $-\frac 1{\sqrt 2}< \tau < \frac 1{\sqrt 2}$, and Euclidean for  $\frac 1{\sqrt 2}< | \tau |<1$.  The latitudes with 
 $ | \tau |=\frac 1{\sqrt 2}$ produce a singularity in the Ricci scalar, and  unlike with the curvature of a Euclidean sphere, the Ricci scalar is not constant and it is negative
\be {\tt R}=-\frac 2{(1- 2\tau^2)^2}\ee
The solution  represents a closed space-time cosmology with the initial and final singularity occurring at the latitudes with 
 $ | \tau |=\frac 1{\sqrt 2}$, where the spatial radius is not zero. The solution can be expressed in terms of the $su(2)$ Poisson bracket algebra. 

\end{enumerate}

\noindent
 The Poisson brackets of the solutions define three-dimensional Lie algebras and it is  straightforward to find their matrix analogues.   One gets  noncommutative de Sitter space-time for both solutions 1. and 2., which we discuss in the next subsection, while for 3.  the result is a Lorentzian fuzzy sphere.\cite{Chaney:2015mfa}.
We note that there are also no  cylindrical space-time solutions, $a=$constant, when $\omega\ne0$.

In general, solutions to (\ref{3difeqspctwne0}) are labeled by four independent parameters: $\upsilon$, $\omega$,  $c_1$ [the integration constant in (\ref{dynfnfrh})] and the value  of $a'$ at some $\tau=\tau_0$.  (The value of $a$ at a given $\tau$ merely determines the overall scale.)   The solutions can be obtained numerically and some novel  results  appear when $\omega\ne 0$. In figure 4 we plot $\tau$ versus $a(\tau)$ for $\upsilon=0$ and different values of $\omega$.  A closed space-time results for  $\omega\le 1$, a dumbbell shaped curve [with two  maxima for $a(\tau)$] appears for $1<\omega<2$ and an open space-time for  $\omega\ge 2$.
 Another example, depicted in figure 5,  shows that starting from initial conditions associated with a rapid inflationary period, one can get a smooth transition to a non inflationary phase. There we take  $a=1$ and $a'=50$ at some initial time  $\tau_0=.01$, along with  the following choices for the remaining parameters: $\upsilon=1,\;\omega=-.5$ and $c_1=-2500$.
We then find that starting with large values for the expansion and deacceleration rates, $a'(\tau)/a(\tau)$ and $-a''(\tau)/a(\tau)$, respectively, at $ \tau=\tau_0$ these quantities goes rapidly to zero as $\tau$ increases  from the initial value.
\begin{figure}[placement h]
\begin{center}
\includegraphics[height=3.75in,width=3.1in,angle=0]{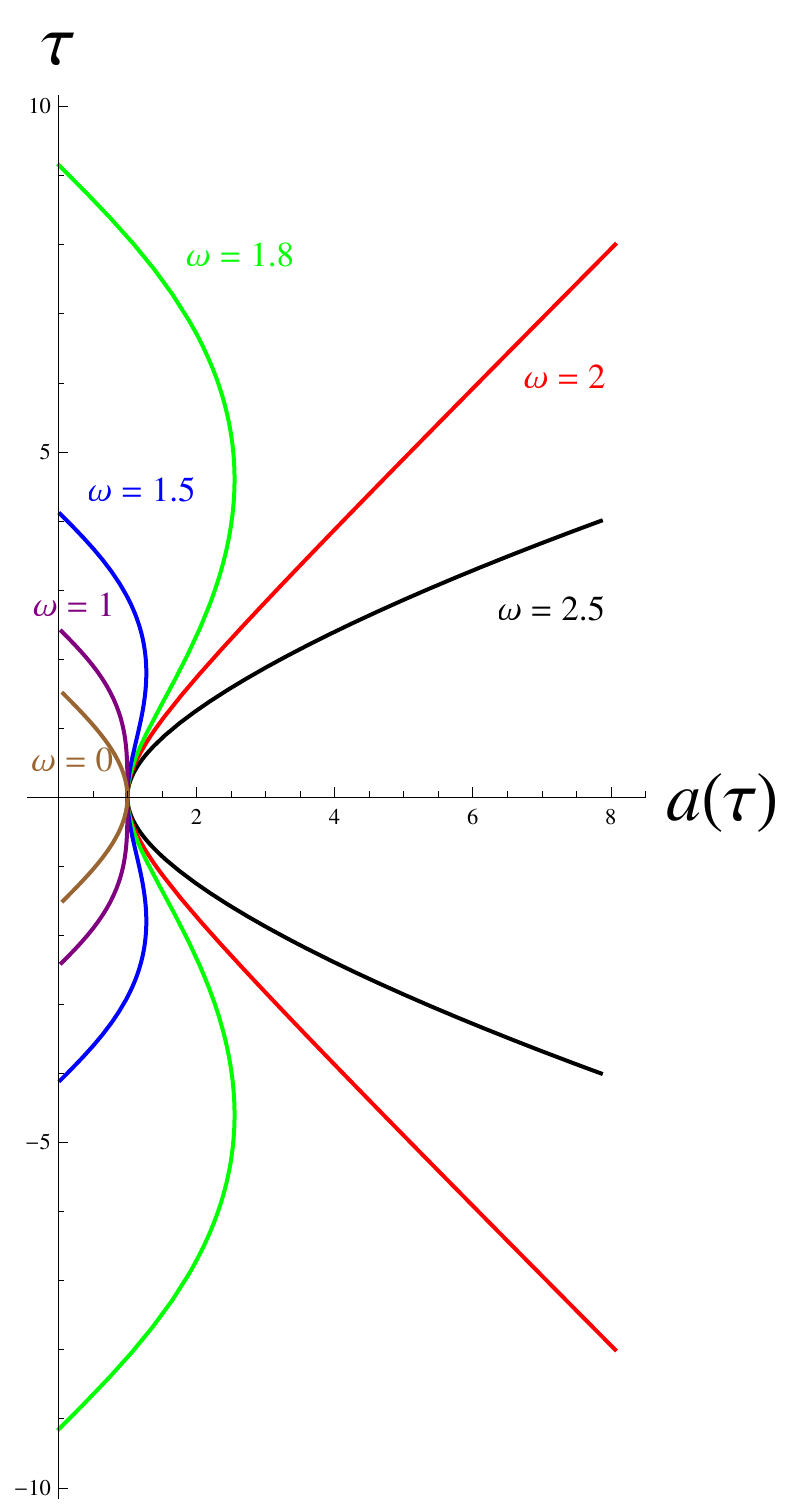}
\caption {Numerical solutions to (\ref{3difeqspctwne0}) for $ \upsilon=0  $ and $\omega= 1,\; 1.5,\; 1.8,\; 2$ and  $2.5$. The boundary conditions are $a(0)=1$ and $a'(0)=0$.
}
\end{center}
\end{figure}
\begin{figure}[placement h]
\begin{center}
\includegraphics[height=3.in,width=3in,angle=0]{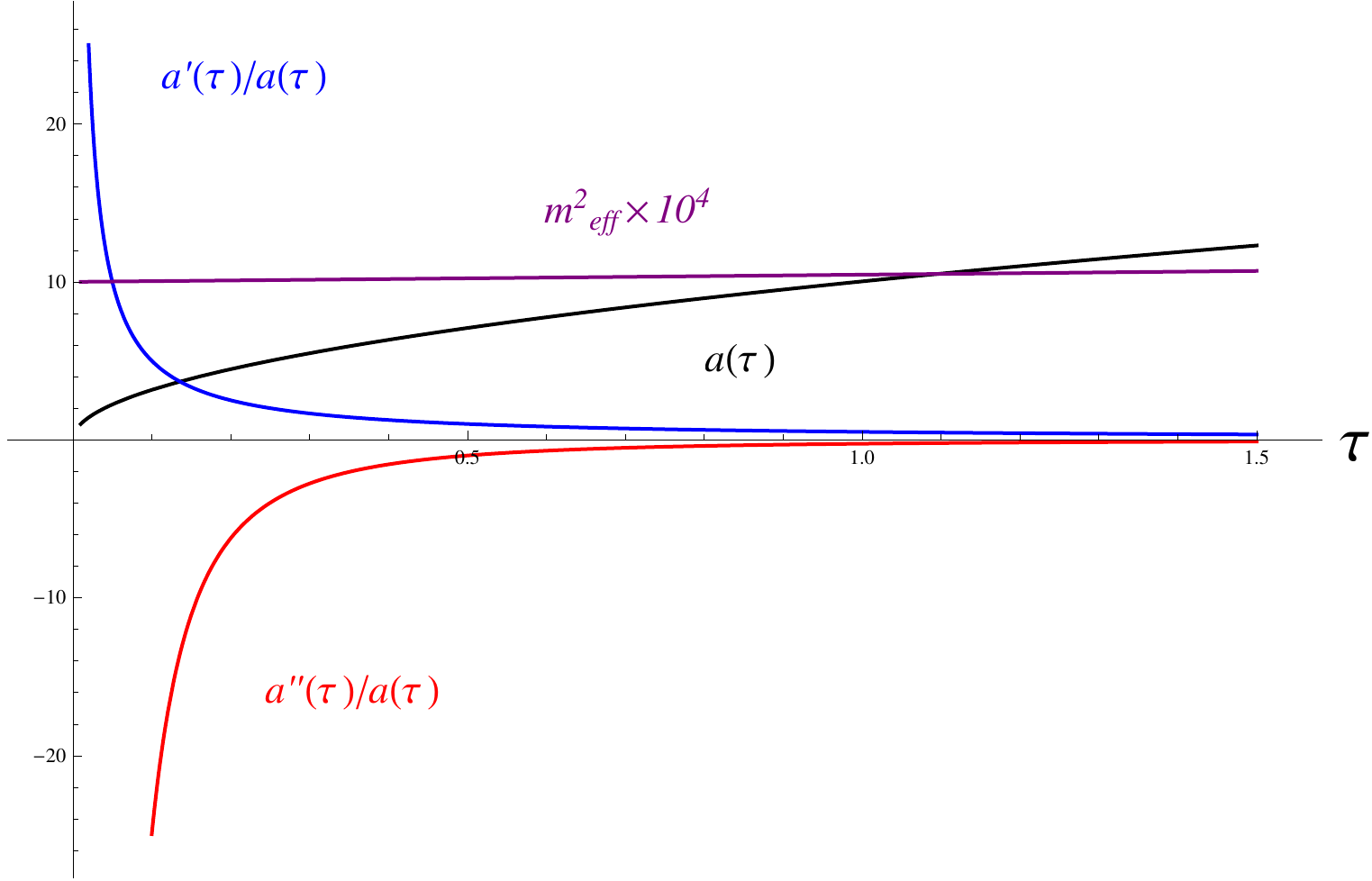}
\caption {Numerical solution for $\upsilon=1,\;\omega=-.5$ and $c_1=-2500$, with initial conditions $a(.01)=1$ and $a'(.01)=50$. Plots are given for $a(\tau)$, $a'(\tau)/a(\tau)$ and $a''(\tau)/a(\tau)$, and show that $a'(\tau)/a(\tau)$ and $-a''(\tau)/a(\tau)$ decreases rapidly to zero as $\tau$ increases  from the initial value $\tau=.01$. We also plot the effective mass-squared $m^2_{eff}$ of the scalar field, which is associated with perturbations normal to the surface. [See subsection 4.2.] The plot indicates that  $m^2_{eff}>0$.
}
\end{center}
\end{figure}

\subsection{Rotationally invariant matrix solutions}
\subsubsection{Recursion relations}
When $\beta\ne 0$, the right hand sides of the matrix equations of motion (\ref{eominAH}) no longer vanish and
substitution of the  ansatz (\ref{WAeit}) into the equations of motion (\ref{eomwthqdtrm}) yields
\beqa \Bigl( X^0(\hat t)-X^0(\hat t-\Delta)-\tilde \alpha\Bigr)A(\hat t)^2
- \Bigl( X^0(\hat t+\Delta)-X^0(\hat t)-\tilde\alpha\Bigr)A(\hat t+\Delta)^2 +\beta X^0(\hat t)=0
&&\cr&&\cr
\frac 12 \Bigl( A(\hat t-\Delta)^2+A(\hat t+\Delta)^2-2A(\hat t)^2 \Bigr)+\Bigl( X^0(\hat t)-X^0(\hat t-\Delta)-\tilde\alpha\Bigr)^2-\tilde\alpha^2-\beta=0&&
\label{useBCHbne0}
\eeqa
So unlike in subsection 2.2.1, $\Bigl( X^0(\hat t)-X^0(\hat t-\Delta)-\tilde\alpha\Bigr)A(\hat t)^2$ is not invariant under discrete translations $\hat t\rightarrow \hat t+ n\Delta$, $n=$integer, except for $\beta = 0$.
The recursion relations (\ref{rcrsnrltnah}) for the eigenvalues (\ref{x0nan}) are now generalized to
\beqa \Bigl( x^0_n- x^0_{n+1}-\tilde\alpha\Bigr)a_n^2
- \Bigl(x^0_{n-1}-x^0_n-\tilde\alpha\Bigr)a_{n-1}^2 +\beta x^0_n=0
&&\cr&&\cr
\frac 12 \Bigl(a_{n+1}^2+a_{n-1}^2-2a_n^2 \Bigr)+\Bigl(x^0_n-x^0_{n+1}-\tilde\alpha\Bigr)^2-\tilde\alpha^2-\beta=0&&\label{rcrsnrltnahbne0}
\eeqa

Once again, starting with any two neighboring eigenvalues for $A^2$ and one time eigenvalue $x^0_n$ one can use the recursion relations to generate a matrix solution.  As before solutions are only valid providing all $a_n^2\ge 0$.  This means that  either $n$ spans all positive and negative integers or the series is terminated at some $n$ [and  then (\ref{WAeit}) no longer holds].  The latter was the case for the discrete series of the noncommutative de Sitter solution. We discuss  noncommutative de Sitter solutions of the equations of motion   (\ref{eomwthqdtrm})  in  subsection 3.2.2.  For finite dimensional matrix solutions there must be both a largest and smallest value of $n$.  In subsection 3.2.3 we show that finite matrix solutions exist for $\beta\ne 0$, and they contain the Lorentzian fuzzy sphere.\cite{Chaney:2015mfa}

\subsubsection{Noncommutative $dS^2$}
When $\beta\ne 0$, and $\tilde\alpha^2+2\beta>0$, there exist two distinct
solutions to the recursion relations (\ref{rcrsnrltnahbne0}) of the form (\ref{ncds2ltc}), i.e.,
\be x^0_{n}= -\alpha_\pm( n + \epsilon_0) \qquad\qquad a_{n}^2=\alpha_\pm^2n( n +2\epsilon_0+1 )+ a_0^2\label{ncds2ltcbne0}\;,\ee
where
\be \alpha_\pm=\frac 1 2(\tilde\alpha\pm \sqrt{\tilde\alpha^2+2\beta} \,) \label{2sdsolns}\ee
They are associated with two distinct noncommutative de Sitter solutions.  In the commutative limit ($ \tilde\alpha\rightarrow \upsilon \Theta$, $ \beta\rightarrow \omega \Theta^2$) they  go to the two de Sitter space-times described in subsection 3.1, with $\alpha_\pm\rightarrow \Theta h_\pm$ and $ h_\pm$ given in (\ref{2dstnctds2}). In both cases $X^\mu$ span an $su(1,1)$ Lie algebra (\ref{suoneone}) with noncommutative parameters $\alpha=\alpha_\pm$. The two noncommutative solutions coincide when $\tilde\alpha^2+2\beta=0$.  In the limit $\beta\rightarrow 0$,
one of them ($\alpha_+\rightarrow\tilde\alpha$) reduces to the solution of  subsection 2.24, while the other ($\alpha_-\rightarrow 0$) becomes the vacuum solution. Once again,  nontrivial representations fall into the three categories, principal, supplementary and discrete series, and they can be constructed as in subsection 2.2.4 for each of the two solutions.

The noncommutative  $dS^2$ relations (\ref{suoneone}) also solves the matrix  equations (\ref{eomwthqdtrm}) in the limit $\beta,\;\tilde\alpha\rightarrow\infty$, with $\frac\beta{\tilde\alpha}$ finite.   This corresponds to a matrix action where the kinetic energy (or Yang-Mills) term is absent. The noncommutative  $dS^2$ solution in this case has
\be\alpha=\frac\beta{2\tilde\alpha}\;,\ee
and is the matrix analogue of (\ref{2ddsnke}).

\subsubsection{Finite dimensional matrix solutions}
		When $\beta\ne 0$, there exist finite dimensional matrix solutions of the equations of motion (\ref{eomwthqdtrm}), which are associated with the $su(2)$ algebra. For this we express the matrices $X^\mu$ as a linear combination of $J_i,$ which are $N\times N$ hermitean matrices spanning the $su(2)$ algebra, $[J_i,J_j]=i\epsilon_{ijk} J_k$. Here $i=1,2,3$ are {\it  Euclidean} indices.  We consider the following linear combination
\be
X^0=2 J_3\qquad X^1=2v_1 J_1\qquad X^2=2( v_2 J_2+ v_3 J_3) \;\ee
$X^\mu$ are a solution to the equations of motion (\ref{eomwthqdtrm}) when
\be\beta = -4 v_1^2\qquad\quad \tilde\alpha^2= 1 -\frac{v_3^2}{1+v_1^2}\qquad\quad v_2= -\tilde\alpha v_1\ee
Hermiticity  requires all coefficients $v_i$ to be real and thus  $\beta<0$ and $\tilde\alpha^2\le 1$. Given
that $J_i$ define an irreducible representation of $su(2)$, the time eigenvalues for this solution are
\be 2m = -N+1,-N+3,..., N-1\ee
Like with the fuzzy sphere embedded in three-dimensional Euclidean space,\cite{Madore:1991bw}-\cite{Iso:2001mg} the matrices $X^\mu$  span the $su(2)$ algebra. Here the  $su(2)$ generators are given by
\be J_1=\frac 1{2 v_1} X^1\qquad\quad J_2=\frac 1{2 v_2} ( X^2-v_3 X^0)\qquad\quad J_3=\frac 1{2 } X^0\;, \ee for $v_1,v_2\ne 0$
and $X^\mu$ solve the  Lorentzian, rather than Euclidean, matrix equations.

The above solutions are not in general rotationally invariant; i.e., (\ref{cmxpxmwx0}) may not  apply for these solutions, and therefore neither the ansatz (\ref{WAeit}) nor (\ref{WAU}) in general hold. An exceptional case is $v_3=0$.
After setting $ \beta = -4\tilde\alpha^2 v_1^2$ and doing a rescaling we can write \be
X^0=2 \tilde\alpha J_3\qquad X^1=\sqrt{-\beta}\, J_1\qquad X^2=-\sqrt{-\beta}\, J_2\;,\qquad\;\;\label{rtnvfntsln}\ee
For this case, $X_+X_-$ commutes with $X^0$. We can then identify the matrix $A^2$ with $-\beta (\vec J^2-J_3^2-J_3)$.  The $N-$dimensional irreducible representations for $X^\mu$ satisfying (\ref{rtnvfntsln}) define  Lorentzian fuzzy spheres and were discussed previously in \cite{Chaney:2015mfa}.
The  commutative solution  is recovered by taking $N\rightarrow \infty$ limit,  along with $\tilde\alpha,\beta\rightarrow 0$.  For this we need to keep both $\tilde\alpha N$ and $\sqrt{-\beta }\,N$ finite, with $ \frac{\sqrt{-\beta}}{2\tilde\alpha}\rightarrow a_0\label{eighttwo}$, in the limit.

\section{Stability analysis}

\setcounter{equation}{0}
Here we examine small perturbations about the rotationally invariant solutions obtained in the previous two sections.  After  substituting the perturbations back into the matrix action and taking the commutative limit we obtain a scalar field  coupled to a gauge field on the space-time manifold associated with the commutative solution.  Upon eliminating the gauge fields one gets an effective mass term for the scalar.  As stated earlier,  the effective mass-squared is negative for the systems studied in section 2, as was shown in \cite{Stern:2014aqa}.  The result changes when the quadratic term is included in the matrix model action.  We show that  the effective mass-squared  is positive for a range of $\omega\ne 0$ ensuring the stability of the  field theory in the commutative limit.

 We begin  in subsection  4.1 with the specific example of noncommutative de Sitter space, which is a solution of the matrix model, with or without the inclusion of the quadratic term. [Cf. subsections 2.2.4  and 3.2.2.]  A general analysis which is valid for all rotationally invariant  solutions is carried out  in  subsection  4.2.  It relies on finding the appropriate Seiberg-Witten map for the system.

\subsection{Noncommutative $dS^2$}
Here we expand the matrices $Y^\mu$ about the noncommutative de Sitter solution $X^\mu$, which satisfy the $su(1,1)$ commutation relations (\ref{suoneone}).  Taking the expansion parameter to be the noncommutative parameter $\alpha$, one can write
\be Y^\mu= X^\mu {\color{black}-} \alpha R\, A^\mu\;,\label{dsxpnsn}\ee
at leading order.  $R$ is a distance scale, with
 $R^2$ being the value of the central operator in (\ref{copsu11}).  As in \cite{Iso:2001mg}, the infinite dimensional   hermitean  matrices $A^\mu$ are functions of $ X^\mu$. $A^\mu$  can be regarded as noncommutative potentials.  This is since infinitesimal gauge transformations iii) of the form $U=\BI {\color{black}+}i\alpha R\Lambda$, where $\Lambda$ is an infinite dimensional hermitean matrix with infinitesimal elements, lead to the gauge variations
\be \delta A^\mu =-i [\Lambda , X^\mu] {\color{black}+}i
\alpha R [\Lambda , A^\mu]\ee
Following \cite{Iso:2001mg} we define noncommutative field strengths $F_{\mu\nu}$ according to
\be
\alpha^2 R^2 F_{\mu\nu}=[Y_\mu,Y_\nu]  {\color{black}+}i\alpha \epsilon_{\mu\nu\lambda} Y^\lambda \label{Fmunu}\; \ee They transform covariantly, $\delta F_{\mu\nu} = {\color{black}+}i\alpha R [\Lambda , F_{\mu\nu}]$.  $F_{\mu\nu}$  vanishes when it is evaluated on the noncommutative de Sitter solution $Y^\mu=R X^\mu$, where $X^\mu$ satisfies (\ref{suoneone}).

We next express the matrix model action in terms of $A_\mu$ and $F_{\mu\nu}$.  We do this  for the action (\ref{mmactn}) in subsection 4.1.1 and then consider the quadratic term in 4.1.2.

\subsubsection{$\beta=0$}
Upon substituting (\ref{dsxpnsn})  and (\ref{Fmunu})   into the action (\ref{mmactn}) we get
\be S(Y)=\frac {\alpha^4R^2}{g^2}{\rm Tr}\Bigl\{ -\frac{R^2}4 F_{\mu\nu} F^{\mu\nu}
-\frac {iR}6 \epsilon_{\mu\nu\lambda} F^{\mu\nu}A^\lambda
 {\color{black}+}\frac i{6\alpha} \epsilon_{\mu\nu\lambda} [A^\mu, X^\nu] A^\lambda +\frac 16 A_\mu A^\mu\Bigr\}\; +\; S( X)\label{four.4}\;
\ee
The result can be re-expressed on the two-dimensional de Sitter manifold, with embedding coordinates $x^\mu$ satisfying
$ ( x^1)^2+ ( x^2)^2-(x^0)^2=R^2$,  upon using  the appropriate star product for de Sitter space\cite{Jurman:2013ota}. The latter can be expanded in  the noncommutative parameter $\alpha$.  At lowest order, the star commutator of the symbols of the functions ${\cal F} ( X)$ and ${\cal G} ( X)$ goes to $ {\color{black}+}i\alpha/R$ times the Poisson bracket of ${\cal F} (x)$ and ${\cal G}(x)$, i.e., $ {\color{black}+}\frac{ i\alpha}R\{{\cal F}(x),{\cal G}(x)\}$. Then the commutative limit of the action (\ref{four.4}) may be written as an integral of symbols according to
\be S(X)-S(\bar X)\rightarrow \frac{\alpha^4 R^2}{g_c^2}\int d\mu \;\Bigl\{ \frac 14 \Bigl(\{A_\mu,x_\nu\}-\{A_\nu,x_\mu\}\Bigr)^2 -\frac R2\epsilon_{\mu\nu\lambda} \{A^\mu,x^\nu\}A^\lambda\Bigr\}\label{0.9}\;,\ee
where $d\mu$ is the invariant measure over de Sitter space which we now specify.
Instead of using (\ref{ncprtztn}),
we find it more convenient to use a different parametrization of the the surface. Following \cite{Jurman:2013ota}, we take
\be \pmatrix{x^0\cr x^1\cr x^2}=R \pmatrix{\tan \eta \cr \sec\eta\cos \sigma\cr
\sec\eta \sin\sigma
}\;,\label{adstwo}\ee where $\{-\frac\pi 2\le \eta\le\frac\pi 2\,, -\pi\le \sigma<\pi\} $ and we have included the distance scale. For the integration measure we take $d\mu= d\eta d\sigma\, R^2/\cos^2\eta$.
The induced metric tensor resulting from (\ref{adstwo}) is
\be - {\tt g}_{\eta\eta}= {\tt g}_{\sigma\sigma}=\frac {R^2}{\cos^2\eta}\qquad\quad {\tt g}_{\eta\sigma}=0\;\label{ds2mtrc}\ee The fundamental Poisson brackets are $\{\eta,\sigma\}=\cos^2\eta$, which is consistent with the $su(1,1)$ Lie algebra,  $\{x^\mu,x^\nu\}=R\epsilon^{\mu\nu\lambda}x_\lambda $.

Like in \cite{Iso:2001mg}, we introduce a pair of tangent vectors $K^a_\mu$, $a=\eta,\sigma$ on the manifold
defined by
\be K^a_\mu=\frac{\cos^2\eta}{R^2}\epsilon^{ab}\epsilon_{\mu\nu\lambda}\, x^\nu\partial_bx^\lambda\;\ee $K^a_\mu$ along with the normal vector $x_\mu$ form the orthogonal basis.
Moreover,
\be K_\mu^a K^{b\mu} =R^2 {\tt g}^{ab} \qquad K_\mu^\eta x^{\mu}=K_\mu^\sigma x^{\mu}=0\qquad \epsilon_{\mu\nu\lambda} K^{\eta\mu}K^{\sigma\nu}x^\lambda =R\cos^2\eta\ee
Additional identities are:\beqa &&\frac{\partial K^\eta_\mu}{\partial \eta}=\frac{\partial K^\sigma_\mu}{\partial \sigma}=-\tan\eta K^\eta_\mu -\frac{1}{R }x_\mu \qquad\quad \frac{\partial K^\sigma_\mu}{\partial \eta}= \frac{\partial K^\eta_\mu}{\partial \sigma}=-\tan\eta K^\sigma_\mu\cr &&\cr
&& \frac{\partial x_\mu}{\partial \eta}=-\frac R{\cos^2\eta} K^\eta_\mu\qquad\qquad\qquad\qquad \qquad\frac{\partial x_\mu}{\partial \sigma}=\frac R{\cos^2\eta} K^\sigma_\mu \;\eeqa
We now expand $A_\mu(\eta,\sigma)$ in the above basis and define $U(1)$ gauge potentials $({\cal A}_\eta,{\cal A}_\sigma)$ and a scalar field $\phi$ on de Sitter space: 
\be R A_\mu(\eta,\sigma) = {\cal A}_\sigma(\eta,\sigma) K^\eta_\mu+ {\cal A}_\eta(\eta,\sigma) K^\sigma_\mu+ \frac 1R \phi(\eta,\sigma)\, x_\mu\label{ramu} \ee
Then using the above identities
\beqa \{A_\mu,x_\nu\}-\{A_\nu,x_\mu\}&=& {\cal F}_{\eta\sigma}K^\eta_\mu K^\sigma_\nu+\phi \epsilon_{\mu\nu\lambda} \frac{x^\lambda}R \qquad\cr&&\cr
&&+( \partial_\eta \phi-{\cal A}_\sigma)\frac{x_\mu}R K^\sigma_\nu +(\partial_\sigma \phi-{\cal A}_\eta)\frac{x_\mu}R K^\eta_\nu\quad -\quad(\mu \rightleftharpoons \nu)\;,\qquad\label{damunu}\eeqa
where ${\cal F}_{\eta\sigma}=\partial_\eta {\cal A}_\sigma-\partial_\sigma {\cal A}_\eta$ is the $U(1)$ field strength on the de Sitter manifold.
Substituting (\ref{ramu}) and (\ref{damunu}) into
the commutative limit (\ref{0.9}) of the action gives
\beqa S(X)-S(\bar X)&\rightarrow &\frac{\alpha^4 R^2}{g_c^2}\int {d\eta d\sigma} \;\Bigl\{-\frac 12 \cos^2\eta\,({\cal F}_{\eta\sigma})^2+\frac 12 (\partial_{\eta} \phi)^2 -\frac 12 (\partial_{\sigma} \phi)^2+2 \phi {\cal F}_{\eta\sigma}-\frac {\phi^2}{\cos^2\eta}\Bigr\}\label{eighteight}\cr &&\cr &\rightarrow &\frac{\alpha^4 R^2}{g_c^2}\int {d\eta d\sigma}\sqrt{-{\tt g}} \;\Bigl\{\frac{R^2}4 \,{\cal F}_{ab}{\cal F}^{ab} -\frac 12 \partial_{a} \phi\partial^a\phi+\frac 2{\sqrt{-{\tt g}}} \phi{\cal F}_{\eta\sigma}-\frac 1{R^2}{\phi^2}\Bigr\}\;,\label{fourone3}\eeqa
where ${\tt g}$ is the determinant of the induced metric (\ref{ds2mtrc}).  The mass-squared for the scalar field appearing in the action  is positive, $m^2=2/R^2$.  However, the gauge field is nondynamical on a two-dimensional space-time and can be eliminated, which leads to the effective field Lagrangian
\be {\cal L}_{\rm eff}\;= \;-\frac 12 \partial_{a} \phi\partial^a\phi+\frac 1{R^2}{\phi^2}\;\label{fctvl}\ee  We then get a switch of  sign for the mass-squared of the scalar field, and so the scalar field is tachyonic.  More generally,  tachyonic excitations were shown to occur for all spherically symmetric solutions of the matrix model described in section 2.\cite{Stern:2014aqa}

We note that the kinetic energies of the gauge and scalar fields  in (\ref{fourone3}) have opposite signs.  This appears to be a generic feature of the Lorentzian matrix model,\cite{Stern:2014aqa},\cite{Chaney:2015mfa}
and is not totally unexpected since the  matrix model action, specifically the Yang-Mills term, is not positive definite.  This situation is harmless in two space-time dimensions since the gauge field can be eliminated.  However, the same does not apply in higher dimensions, and this issue is yet to be resolved.

\subsubsection{$\beta\ne 0$}

We now include the quadratic term in the total action (\ref{mmactnplsqd}). As stated in sec 3.2.2, there exist two noncommutative de Sitter solutions (\ref{suoneone}) when $\tilde\alpha^2+2\beta>0$ (and one when $\tilde\alpha^2+2\beta=0$).  They correspond to $\alpha=\alpha_\pm$ as given in (\ref{2sdsolns}).
There is also one noncommutative de Sitter solution in the limit  $\beta,\;\tilde\alpha\rightarrow\infty$, with $\frac\beta{\tilde\alpha}$ finite. In this case, 
$\alpha= {\color{black}+}\frac\beta{2\tilde\alpha}$.

An expansion (\ref{dsxpnsn}) in the action (\ref{mmactnplsqd}) around one of the
de Sitter solutions
gives
\beqa S_{\tiny {\rm total}}(Y)&=&\frac {\alpha^4R^2}{g^2}{\rm Tr}\Bigl\{ -\frac{R^2}4 F_{\mu\nu} F^{\mu\nu}
- {i\gamma R} \epsilon_{\mu\nu\lambda} F^{\mu\nu}A^\lambda
+\frac {i\gamma}{\alpha}\epsilon_{\mu\nu\lambda} [A^\mu, \bar X^\nu] A^\lambda +\gamma A_\mu A^\mu\Bigr\}\cr &&\cr &&\quad\qquad \; +\; S_{\tiny {\rm total}}( X)\label{0.4}\;,
\eeqa
where $\gamma=\frac 12 {\color{black}-}\frac{\tilde\alpha}{3\alpha}$ and $F_{\mu\nu}$ is again defined by (\ref{Fmunu}). We now repeat the previous procedure to obtain the commutative limit of the action.  $\frac{\tilde\alpha}{\alpha}$ and $ \frac{\beta}{\alpha^2}$ go to finite values in the commutative limit, respectively, ${\color{black}+ }\frac{\upsilon}{h}$ and $ \frac{\omega}{h^2}$, and hence so does $\gamma$.
One now gets
\beqa S_{\tiny {\rm total}}(Y)-S_{\tiny {\rm total}}(X)&\rightarrow& \frac{\alpha^4 R^2}{g_c^2}\int \frac{d\eta d\sigma}{\cos^2\eta} \;\Bigl\{ \frac 14 \Bigl(\{A_\mu,x_\nu\}-\{A_\nu,x_\mu\}\Bigr)^2\cr &&\cr &&\quad\qquad \;+\Bigl(\frac12{\color{black}- }\frac {\upsilon}h\Bigr)R\epsilon_{\mu\nu\lambda} \{A^\mu,x^\nu\}A^\lambda+\Bigl(1{\color{black}-}\frac {\upsilon}h\Bigr)R^2A_\mu A^\mu\Bigr\}
\cr &&\cr &&\cr
&\rightarrow &\frac{\alpha^4 R^2}{g_c^2}\int {d\eta d\sigma}\sqrt{-{\tt g}} \;\Biggl\{\frac{R^2}4 \,{\cal F}_{ab}{\cal F}^{ab} -\frac 12 \partial_{a} \phi\partial^a\phi\cr &&\cr &&\qquad\qquad\qquad\quad+\;\Bigl(2{\color{black}- }\frac{\upsilon}h\Bigr)\biggl(\frac 2{\sqrt{-{\tt g}}} \phi {\cal F}_{\eta\sigma}-\frac 1{R^2}{\phi^2}\biggr)\Biggr\}\;,\label{four1six}
\eeqa
in the commutative limit after applying (\ref{ramu}) and (\ref{damunu}). The  kinetic energy terms are unaffected by the deformation parameter $\beta$, and  the previous result (\ref{eighteight}) for the remaining terms are recovered when $h\rightarrow{\color{black}+ }\upsilon$ corresponding to $\omega\rightarrow 0$.
Upon eliminating the nondynamical gauge field, one gets the following effective field Lagrangian
\be {\cal L}_{\rm eff}\;= \;-\frac 12 \partial_{a} \phi\partial^a\phi+\frac 1{R^2} \Bigl(2{\color{black}- }\frac{\upsilon}h\Bigr)
\Bigl(3{\color{black}- }2\frac{\upsilon}h\Bigr){\phi^2}\;,\label{effLdSttr}\ee
generalizing (\ref{fctvl}).
Thus the mass-squared now depends on $\frac{\upsilon}h$, which from (\ref{2dstnctds2}) has two possible values, $\frac{\upsilon^2}\omega(-1\pm\sqrt{1+\frac{2\omega}{\upsilon^2}})$.  It is positive for
\be \frac32 <{\color{black}+ }\frac{\upsilon}h<2\;,\ee and so the system is stabilized for this range of parameters. The value $\omega=0$ corresponds to ${\upsilon}/h\rightarrow{\color{black}+ }1$  or $\infty$  and hence it
is not included in this range.  This is consistent with the result of the previous subsection that the mass-squared is negative when the quadratic term is not included in the matrix model. The corresponding values for $\omega$ are obtained using $\omega=2h(h{\color{black}- }\upsilon)$.
The scalar field is massless for the special values
$(\frac{\upsilon}h,\frac\omega{2h^2})=({\color{black}+ }\frac 32, -\frac 12)$ and $({\color{black}+ }2, - 1)$.

\subsection{General analysis using a Seiberg-Witten map}
Here we consider perturbing about arbitrary solutions to the matrix equations (\ref{mmactnplsqd}). Since in general we don't have exact solutions, we cannot follow the previous procedure, which requires defining a noncommutative field strength that vanishes for zero perturbations. An alternative procedure is to expand the action about the commutative solution up to second order in the noncommutative parameter $\Theta$. The perturbations can be expressed in terms of  commutative gauge fields and a scalar field upon applying the appropriate  Seiberg-Witten map on the two-dimensional target manifold  ${\cal M}_2$. The map was found in \cite{Stern:2014aqa} and is reviewed in Appendix A.  The commutative gauge fields correspond to perturbations along the tangent directions of ${\cal M}_2$, while the scalar field is associated with perturbations normal to the surface. In \cite{Stern:2014aqa}, the procedure was applied to the rotationally invariant manifolds which were solutions to the equations of motion (\ref{clmeas}).  These equations followed from the action (\ref{cmtvlmtsc}) which did not include a quadratic term. There we found that for all rotationally invariant solutions, the effective dynamics for the scalar field gave a tachyonic mass. This indicates an instability with respect to perturbations normal to the surface. Here we include the quadratic term in the matrix model action,  i.e., we start with 
(\ref{mmactnplsqd}), and
repeat the stability analysis for the rotationally invariant solutions to equations of motion (\ref{clmeaswne0}). Just as in  subsection 4.1.2, we find that when $\beta\ne 0$, the effective mass-squared for the scalar field can be positive for a certain range of parameters.

As in 4.1.1 and 4.1.2, we  denote perturbations of the embedding coordinates by $A^\mu$. They are perturbations about the commutative space-time solutions, as well as, functions on the noncommutative manifold. Assuming the existence of a general rotationally invariant star product one can write the perturbations as functions on a smooth manifold parametrized  by $\tau$ and $e^{i\sigma}$ , and so $A^\mu=A^\mu(\tau,e^{i\sigma})$. The perturbation parameter shall again be identified with the noncommutativity parameter $\Theta$, and thus
\be y^\mu=x^\mu+ \Theta A^\mu \;\label{yxbta}\ee
The perturbations (\ref{yxbta}) induce nonvanishing fluctuations in the induced metric tensor at first order in $\Theta$, and thus  affect the space-time geometry. As in the previous subsection, $A_\mu$ transform as noncommutative gauge potentials. Up to first order in $\Theta$, the infinitesimal gauge variations of $A_\mu$ are given by
\be \delta A_\mu=\{\Lambda, x_\mu\} +\Theta\{\Lambda, A_\mu\}\label{gvrtnsAmu}\ee
Using the Poisson brackets (\ref{pbtauphi}), gauge variations at zeroth order in $\Theta$ are along the tangential directions of ${\cal M}_2$,
\be\delta A_\mu=- h\Bigl(\partial_\tau\Lambda\partial_\sigma x_\mu-\partial_\sigma\Lambda\partial_\tau x_\mu\Bigr)+{\cal O}(\Theta)\ee
 We shall choose $h=h(\tau)$, which is consistent with all the solutions in subsections 2.1 and 3.1.

Using a Seiberg-Witten map\cite{Seiberg:1999vs}, the noncommutative potentials $A_\mu$ can be re-expressed in terms of commutative gauge potentials, denoted by $({\cal A}_\tau,{\cal A}_\sigma)$, on ${\cal M}_2$, along with their derivatives. Since the noncommutative potentials $A_\mu$ have three components and the commutative potentials have only two, an additional degree of freedom, associated with a scalar field $\phi$ should be included in the map: $A_\mu=A_\mu[{\cal A}_\tau,{\cal A}_\sigma,\phi]$. The Seiberg-Witten map is defined so that commutative gauge transformation, $({\cal A}_\tau,{\cal A}_\sigma)\rightarrow ({\cal A}_\tau+\partial_\tau\lambda,{\cal A}_\sigma+\partial_\sigma\lambda)$, for arbitrary functions $\lambda$ on ${\cal M}_2$,  induce noncommutative gauge transformations for $A_\mu$, which are given by (\ref{gvrtnsAmu}) for infinitesimal gauge transformations. $\Lambda$ in this case is a function of $\lambda$, along with commutative potentials and their derivatives, $\Lambda=\Lambda[\lambda,{\cal A}_\tau,{\cal A}_\sigma]$.

The Seiberg-Witten map consistent with (\ref{pbtauphi}) and (\ref{ncprtztn}) was obtained in \cite{Stern:2014aqa} up to first order
in $\Theta$ and is written down explicitly  in Appendix A, c.f., (\ref{axpnnbta})-(\ref{swone}). In this regard the first order map is sufficient for our purposes
since we wish to expand the action $ S_c$, and hence also $y^\mu$, up to second order in $\Theta$.  The task is to thus  substitute (\ref{yxbta}), along with the map (\ref{axpnnbta})-(\ref{swone}) into the action (\ref{cmtvlmtscbne0}). After some work we find
\be
S_c(y)=-\frac {\Theta^2}{g_c^2}\int d\tau d\sigma\,h(\tau)^3{\tt g}^2\,\biggl(\frac 12 {\cal F}_{\tau\sigma}{\cal F}^{\tau\sigma}-\frac 12\partial_{\tt a}\phi\partial^{\tt a}\phi+\gamma(\tau){\cal F}_{\tau\sigma}\phi-\frac 12 m^2(\tau)\phi^2\biggr)\;+\; S_c(x)\;,\label{251}
\ee
where indices ${\tt a,b,...}=\tau,\sigma$ are raised and lowered with the induced metric associated with the invariant interval (\ref{gnrl2dia}). The time-dependent coupling coefficient $\gamma(\tau)$ and mass $m(\tau)$ are given by
\beqa \gamma(\tau)&=&
\frac{2 a(\tau)^2}{h(\tau)^2{\tt g}^2}\Bigl({\color{black}- }\upsilon h(\tau){\tt g}_{\tau\tau}+\omega a(\tau)\Bigl(\frac\tau{a(\tau)}\Bigr)' \Bigr)\cr &&\cr
m(\tau)^2&=&
\frac{ a(\tau)^2}{h(\tau)^2{\tt g}^2}\biggl({\tt g}_{\tau\tau}\Bigl( 2 h(\tau)^2{\color{black}- }4\upsilon h(\tau)-\omega\Bigr)+ 2\omega a(\tau) \Bigl(\frac\tau{a(\tau)}\Bigr)'  \biggr)\label{fmamsq}
\eeqa
Here we see a common feature for these systems, which is  that the kinetic  energies  of the gauge and scalar fields have opposite signs. 
Upon eliminating the gauge field  using its equation of motion, ${\cal F}^{\tau\sigma}+\gamma(\tau)\phi={\rm constant}/(h(\tau)^3{\tt g}^2)$, and substituting back into  (\ref{251}), we now get the effective action
\be
S^{\rm eff}_c(y)=-\frac {\Theta^2}{g_c^2}\int d\tau d\sigma\,h(\tau)^3{\tt g}^2\,\biggl(-\frac 12\partial_{\tt a}\phi\partial^{\tt a}\phi-\frac 12 m _{\rm eff}(\tau)^2\phi^2\biggr)\;+\; S^{\rm eff}_c(x)\;,\label{1hndred}
\ee
where the mass-squared for the scalar field gets modified to \beqa && m _{\rm eff}(\tau)^2\;=\;  m(\tau)^2 +{\tt g} \gamma(\tau)^2  \cr &&\cr &=& \frac{ 2}{h(\tau)^2 {\tt g}}\Biggl(\upsilon^2+(h(\tau){\color{black}- }\upsilon)^2-\frac\omega 2+\frac{ \omega a(\tau) }{{\tt g_{\tau\tau}}^2 h(\tau)^2} \Bigl(\frac\tau{a(\tau)}\Bigr)'\biggl({\tt g}_{\tau\tau}h(\tau)(h(\tau){\color{black}- }4\upsilon)+2\omega a(\tau)\Bigl(\frac\tau{a(\tau)}\Bigr)'\biggr)\Biggr)\label{gnrlmeffsq}
\cr &&\eeqa

The result for  $m_{\rm eff}^2 $ is in general a function of the time parameter $\tau$ and it can be positive, negative or zero.   It is  negative  when $\omega=0$ for all spherically symmetric solutions.  This follows from the Lorentzian signature of the metric, ${\tt g}<0$.  The result agrees with what was found in \cite{Stern:2014aqa}.

We get the following results for the three exact solutions found in section 3.1:

\begin{enumerate}

\item $\gamma$, $m^2$ and $ m_{\rm eff}^2$ are constants for the de Sitter solution  (\ref{dScmtvesln}) and (\ref{2dstnctds2})  with   $\upsilon$ and  $\omega$  finite.    In this case (\ref{fmamsq}) and   (\ref{gnrlmeffsq})  yield 
\be\gamma=m^2= 2\Bigl(2{\color{black}- }\frac\upsilon h\Bigr)\qquad\qquad m_{\rm eff}^2=2\Bigl(2{\color{black}- }\frac\upsilon h\Bigr)\Bigl(-3{\color{black}+ }2\frac\upsilon h\Bigr) \ee   
Of course this case has already been discussed in section 4.1. The result  agrees with the mass-squared appearing in (\ref{effLdSttr}), which as we saw can be positive, negative or zero. (Here we set $R=1$.)

\item
For the $dS^2$ solution  (\ref{2ddsnke}) obtained  in  the limit  $\upsilon$, $\omega\rightarrow\infty$, with $\frac \omega\upsilon$ finite, the result of the perturbations is a $BF$ Lagrangian with no kinetic energy terms,
\be
{\cal L}_{\rm eff}={\cal F}_{\tau\sigma}\phi-\frac 12 \phi^2\label{4twentee7}
\ee
Thus both the gauge field and scalar field are nondynamical, with the equations of motion giving ${\cal F}_{\tau\sigma}=\phi=$constant.
The result for the field theory action  can also be seen from (\ref{four1six}), since  the kinetic terms do not contribute in the limiting case. Of course the result is not surprising since there are no Yang-Mills terms in the matrix action.

\item  For the case of the sphere embedded in Minkowski space-time (\ref{sphLntz}), we get 
\be \gamma=-\frac{1+2\tau^2}{(1-2\tau^2)^2}\qquad\qquad  
m^2=-\frac{3-2\tau^2}{(1-2\tau^2)^2}\qquad\qquad m_{\rm eff}^2=4\frac{1-\tau^2+ 2\tau^4}{(1-2\tau^2)^3} \ee  The latter is negative in the region with Lorentz signature and hence the scalar field is tachyonic.  As alluded to in subsection 3.1, this solution is a special case of a one-parameter family of solutions associated with ellipsoids in Minkowski space-time.  $ m_{\rm eff}^2 $ is positive for some range of the parameter, corresponding to stable dynamical systems.\cite{Chaney:2015mfa}

\end{enumerate}

For  all the  remaining solutions  $ m_{\rm eff}^2$ can be computed numerically.  In the case of the numerical solution shown  in figure 5 depicting the transition from a rapid inflationary to a non inflationary phase, we find  that $ m_{\rm eff}^2>0$  for all $\tau$.  The result for $ m_{\rm eff}^2$ is also plotted in figure 5.  We can then conclude that the solution is stable for leading order perturbations.

\section{The five dimensional matrix model}

\setcounter{equation}{0}
 We now consider five hermitean matrices $Y^\mu$,  with indices $\mu,\nu,..=0,1,2,3,4$, which are raised and lowered with the five dimensional Minkowski space-time metric $\eta=$diag$(-1,1,1,1,1)$.  
In analogy with section 3, we demand that the dynamics for $Y^\mu$ is invariant under $4+1$ Lorentz transformations and  unitary gauge transformations, but not necessarily  translations.  The Yang-Mills and quadratic terms in the matrix action (\ref{mmactnplsqd}) generalize straightforwardly to the five dimensional case.  The totally antisymmetric cubic term  Chern-Simons type in  (\ref{mmactnplsqd}) can be replaced by a fifth order term.  This term has been introduced previously in Eulidean matrix models.\cite{Kimura:2002nq},\cite{Valtancoli:2002sm},\cite{Azuma:2004yg}
 Then  we can write the  five dimensional matrix action  according to
\be S_{\tiny {\rm total}}(Y)=\frac 1{g^2}{\rm Tr}\Bigl(-\frac 14 [Y_\mu, Y_\nu] [Y^\mu,Y^\nu] +\frac 45  \alpha_5 \,\epsilon_{\mu\nu\lambda\rho\sigma}Y^\mu Y^\nu Y^\lambda Y^\rho Y^\sigma+\frac \beta 2Y_\mu Y^\mu\Bigr)\;\label{mmactnplsqd5d}\;,\ee
where $\epsilon_{01234}=1$ and 
$g$, $\alpha_5$ and $\beta$ are  real constants. The resulting equations of motion are
\be [ [Y_\mu,Y_\nu],Y^\nu]+4\tilde\alpha_5 \, \epsilon_{\mu\nu\lambda\rho\sigma}Y^\nu Y^\lambda Y^\rho Y^\sigma  =-\beta Y_\mu\label{eomwthqdtrm5d}\ee
An exact solution to these equations was found for the Euclidean metric and $\beta=0$ and it was called the fuzzy four sphere.\cite{Kimura:2002nq},  \cite{Valtancoli:2002sm},\cite{Azuma:2004yg}
An analogous construction should be possible in Minkowski space-time to obtain a noncommutative four dimensional de Sitter space. Other nontrivial solutions to this model are not known, however  we can show that  large families of solutions exist after taking the commutative limit of this matrix model.  This is done in subsection 5.1.  One of the solutions is four-dimensional de Sitter space, but it differs from the commutative limit of the Lorentzian counterpart of the four-sphere.  We consider perturbations about the de Sittter solution in subsection 5.2.  Finally in subsection 5.3,  we  propose an ansatz for  rotationally invariant solutions to the five-dimensional matrix model.

\subsection{Solutions in the commutative limit}

In order to take the commutative limit, it is convenient to write the fifth order term in the trace  (\ref{mmactnplsqd5d}) using the commutator, $\frac 15 \alpha_5 \, \epsilon_{\mu\nu\lambda\rho\sigma}Y^\mu [Y^\nu, Y^\lambda ][ Y^\rho, Y^\sigma]$.
Then we can easily apply the usual procedure to get the commutative theory.  We replace matrices $Y^\mu$ by  coordinate functions $y^\mu$, now defined on a four dimensional manifold  ${\cal M}_4$, and matrix commutators by  
$i\Theta$ times the corresponding Poisson bracket  on  ${\cal M}_4$.  $\Theta$ once again denotes the noncommutativity parameter.  If all three terms in the action are to survive in the limit we need that $ \alpha_5$ goes to a finite value, $ \alpha_5\rightarrow \upsilon_5$ and, as before,  $ \beta\rightarrow \omega \Theta^2$, with  $\upsilon_5$ and  $ \omega\;\;{\rm finite}$.   The limiting action in that case is
\be
S_c(y)=\frac 1{g_c^2}\int_{{\cal M}_4} d\mu_4\Bigl(\frac 1{4} \{y_\mu, y_\nu\}\{y^\mu,y^\nu\} -\frac {\upsilon_5 }{5}\epsilon_{\mu\nu\lambda\rho\sigma}\,y^\mu\{y^\nu, y^\lambda\}\{y^\rho, y^\sigma\}+\frac \omega 2 y_\mu y^\mu\Bigr)\;,\label{cmtvlmtscbne05d}
\ee
where 
$ d\mu_4$ denotes the  invariant integration measure, on ${\cal M}_4$.
The  equations of motion resulting from variations of $y^\mu$ are 
\be \{\{y_\mu,y_\nu\},y^\nu\}+\upsilon_5 \epsilon_{\mu\nu\lambda\rho\sigma}\{y^\nu,y^\lambda\}\{y^\rho, y^\sigma\} =\omega y_\mu \label{clmeaswne05d}
\;\ee
Infinitesimal gauge variations again have the form $\delta y^\mu=\Theta\{\Lambda,y^\mu\}$, where  $\Lambda$ is an infinitesimal function on ${\cal M}_4$.

We denote  coordinates on a local patch of  ${\cal M}_4$ by $\tau$, $\sigma$ and  $\xi^i$,  $i,j,...=1,2,3$.  $\tau$ is time-like, while $\sigma$ and  $\xi^i$ are space-like, the latter spanning a unit two-sphere $(\xi^1)^2+(\xi^2)^2+(\xi^3)^2=1$.  A rotationally invariant ansatz for  solutions $y^\mu= x^\mu(\tau,\sigma,\xi^i)$ to the equations of motion (\ref{cmtvlmtscbne05d}) is:
\be \pmatrix{x^0\cr x^i\cr x^4}= \pmatrix{\tau\cr a( \tau)\xi^i \sin\sigma\cr a( \tau)\cos\sigma}\,\label{ncprtztn5d}\ee 
This is an obvious generalization of (\ref{ncprtztn}).    The spatial coordinates
$x^1,\;x^2,\;x^3,\;x^4$ span a three-sphere of radius $a(\tau)$ at time slice,
\be  \vec x^2+ ( x^4)^2=a^2(\tau)\;,\label{cmtvcnstrnt5d}\ee
where $ \vec x^2=x^ix^i$,
and the isometry group is $SO(4)$. For this one assumes $0\le \sigma\le \pi $, with $\sigma=0, \pi $ corresponding to the poles.
The invariant interval on the surface is
\be ds^2=-(1-a'(\tau)^2)\,d\tau^2 +a(\tau)^2\,(d\sigma^2+\sin^2\sigma d s_{{S}^2}^2)\label{gnrl2dia}\;,\ee
where $ds_{{ S}^2}^{2}$ is the invariant interval on the two-sphere and $d\sigma^2+\sin^2\sigma d s_{{S}^2}^2$   is the invariant interval on the three-sphere.
The resulting Ricci scalar and Einstein tensor are nonvanishing, indicating the presence of a gravitational source.  They are, respectively,
\be {\tt  R}=\frac{6(1- a'(\tau)^2+ a(\tau)a''(\tau )}{a(\tau)^2\Big(1-a'(\tau)^2\Bigr)^2}\ee
and \beqa {\tt  G}_{\tau\tau}=\frac 3{a(\tau)^2}&\qquad &{\tt  G}_{\sigma\sigma}=-\frac{\Bigl(1- a'(\tau)^2+2 a(\tau)a''(\tau )\Bigr)}{\Big(1-a'(\tau)^2\Bigr)^2}\cr &&\cr {\tt  G}_{\theta\theta}=\sin^2\sigma\;{\tt  G}_{\sigma\sigma}&\qquad &{\tt  G}_{\phi\phi}=\sin^2\sigma\sin^2\theta\;{\tt  G}_{\sigma\sigma}\;,\eeqa
$\theta$ and $\phi$ being the usual spherical coordinates on the two-sphere.

It remains to define a symplectic structure on the four dimensional space-time manifold.   Although there is no nonsingular Poisson structure on the three sphere, we can write Poisson brackets which are consistent with  three dimensional rotation symmetry, i.e., corresponding to rotations among the three spatial coordinates $x^i$.  The fundamental Poisson brackets are
\be \{\sigma,\tau\}=h(\tau,\sigma)\qquad \qquad \{\xi^i,\xi^j\}=\kappa\epsilon_{ijk} \xi^k\;,\label{frdpbs}\ee 
where $\kappa$ is constant. The Jacobi identity is trivially satisfied. Here, unlike for  the two dimensional case discussed in subsections 2.1 and 3.1,  we  allow for $h$  to be a function of $\sigma$  as well as $\tau$.  
This ansatz along with (\ref{ncprtztn5d}) will allow for nontrivial solutions to the equations of motion.

Below  we examine three different families of solutions to the equations of motion (\ref{clmeaswne05d}).  In each case only two out of the  three terms in the action (\ref{cmtvlmtscbne05d}) contribute :

\begin{enumerate}

\item We first consider the limiting case where both $\,\omega,\,\upsilon_5 \rightarrow \infty$, with $\frac \omega{\upsilon_5}$ and $\kappa$ finite and nonvanishing.   In this limit  the first  term in the action  (\ref{cmtvlmtscbne05d})  (i.e., the Yang-Mills term) does not contribute.  We examined the analogous example in the three-dimensional matrix model (case $2$ in subsection $3.1$) where  the Yang-Mills term does not contribute. The equations of motion are
then 
\be \upsilon_5 \epsilon_{\mu\nu\lambda\rho\sigma}\{y^\nu,y^\lambda\}\{y^\rho, y^\sigma\} =\omega y_\mu \label{clmeaswne05dnym}
\;\ee  They are
 solved by $ a(\tau)^2=\tau^2+1$ which defines the four dimensional de Sitter space $dS^4$
\be     - ( x^0)^2+ \vec x^2+ ( x^4)^2=1   \label{onenineone}\ee
The Poisson structure on this space is determined by the two parameters  $\frac \omega{\upsilon_5}$ and $\kappa$.  The solution for $h(\tau,\sigma)$ is
\be h(\tau,\sigma)=-\frac{\omega}{8\kappa\upsilon_5 }\frac{ \csc^2\sigma}{  a(\tau)^2}\;,\label{oneeighteen}\ee  
 From (\ref{oneeighteen}), the Poisson brackets of the embedding coordinates $x^\mu$  are
\beqa  \{x^0,x^i\}=\frac\omega{8\kappa\upsilon_5}\, \frac {x^i x^4}{( \vec x^2)^{\frac 32}}  &\qquad &
\{x^4,x^i\}=\frac\omega{8\kappa\upsilon_5}\, \frac {x^i x^0}{( \vec x^2)^{\frac 32}} \cr  &&\cr
\{x^0,x^4\}=-\frac\omega{8\kappa\upsilon_5}\,\frac 1{\sqrt{\vec x^2}}&\qquad &
\{x^i,x^j\}=\kappa\sqrt{\vec x^2}\,\epsilon_{ijk}x^k\label{ntpsds4}\eeqa
It can be checked that the Poisson bracket relations are consistent with the de Sitter space condition (\ref{onenineone}).
The Poisson brackets  are  invariant under the action of the three-dimensional rotation  group, although they do not preserve all the isometries of de Sitter space.  More specifically, $SO(4,1)$ is broken to $SO(3)\times {\cal L}_2$, where  ${\cal L}_2$ is the two-dimensional Lorentz group.

This solution is  the four dimensional analogue of the previous $dS^2$ solution (\ref{2ddsnke}) to the equations of motion (\ref{clmeaswne0}) in the limit  $\upsilon$, $\omega\rightarrow\infty$, with $\frac \omega\upsilon$ finite.
Unlike the case with $dS^2$, the Poisson brackets of the coordinate functions $x^\mu$ on $dS^4$ are not associated with  a finite dimensional Lie algebra, and so its matrix analogue of the commutative solution is nontrivial.   By changing the background metric to $\eta=$diag$(-1,1,1,1,-1)$  we can obtain a four dimensional anti-de Sitter space and by switching the background metric to  $\eta=$diag$(1,1,1,1,1)$  we recover   a four sphere.  The solutions are given explicitly in  Appendix  B.

\item Here we set $\kappa= \upsilon_5 =0$ and take
 $h=h(\tau)$.  Now  the second  term in the action  (\ref{cmtvlmtscbne05d}) does not contribute to the dynamics.   The Poisson brackets are noninvertable in this case, and the equations of motion trivially reduce to the two dimensional system (\ref{3difeqspct}) with $\upsilon=0$; i.e. $ h(\tau)$ and $a(\tau)$ should satisfy
\be    (aa'h)'h+h^2=\omega\qquad \quad  2h^2 a'a+a^2h'h=\omega \tau\label{4difeqspct}\;\ee 
 There is a one parameter (not including integration constants) family of solutions which can be obtained numerically.  Solutions for different values of $\omega$  were already plotted in figure 4.    We recall that  $\sigma$, $0\le \sigma<2\pi$, was periodic in sections 2-4. Here $\sigma$ parametrizes  the longitudes on the  three-sphere and ranges from $0$ to $\pi$, where  $\sigma=0$ and $\pi$ denote the poles and correspond to coordinate singularities.

\item Finally we consider $ \omega =0 \;,\; \kappa\ne  0$.  Now  the third  term in the action  (\ref{cmtvlmtscbne05d}) does not contribute.   If we set
\be  h(\tau,\sigma)=f(\tau) \sin^2\sigma\;,\ee
then  the solution to (\ref{clmeaswne05d}) with  space-time index $\mu=0$ is
\be f(\tau)=2\kappa \upsilon_5 a(\tau)^2 +\frac{c_1}{a(\tau)^2}\;,\label{foftau}\ee
where $c_1$ is an integration constant.  The remaining equations of motion, $\mu=i,4$, are solved when $a(\tau)$ satisfies
\be a'(\tau)^2=\Bigl(\frac {\kappa a(\tau)}{f(\tau)}\Bigr)^2+1 \label{aprmoftau} \ee 
 This  implies that  $|a'(\tau)|$ cannot be less than one.  So, for instance,   $a(\tau)$ cannot have turning points and there can be no closed space time solutions. Moreover, from (\ref{gnrl2dia}) the induced metric has a Euclidean signature, even though the background metric is Lorentzian.   Solutions of this form  have no two-dimensional analogues.  They simplify in some limiting cases:

\begin{itemize}

\item  In the case    $ \upsilon_5 \rightarrow 0$, one gets
\be   a'(\tau)^2=\frac{\kappa^2}{c_1^2}\,a(\tau)^6+1  \qquad\qquad f(\tau)=\frac {c_1}{a(\tau)^2}\label{oneone5}\;\ee
$a(\tau)$ is then expressible in terms of inverse elliptic integrals.

\item
The limit $\kappa\rightarrow 0$, $c_1\ne 0$, gives a linearly expanding (or contracting) universe
\be a(\tau)=\pm\tau \qquad\qquad f(\tau)=\frac{c_1}{\tau^2}\;,\ee

\item
Another simplifying limit is $c_1\rightarrow 0$, leading to \be a'(\tau)^2=\frac{1}{4\upsilon_5 a(\tau)^2}+1  \qquad\qquad f(\tau)=2\kappa \upsilon_5 a(\tau)^2 \;,\ee which can be easily integrated
\be a(\tau)=\frac12\sqrt{(2\tau+ c_2)^2-\frac 1{\upsilon_5^2}} \label{solcee} \;,\ee
where $c_2$ is an integration constant. This solution describes an open space-time. Here we must restrict the time domain to $|2\tau+c_2|\ge 1/|\upsilon_5|$.  The solution for $a(\tau)$ goes asymptotically  to $\tau$ and it  is singular in the limit  $ \upsilon_5 \rightarrow 0$, as well as $\kappa\rightarrow 0$.   
\end{itemize}

In general, solutions of (\ref{foftau}) and (\ref{aprmoftau}) are parametrized by $\kappa$,  $\upsilon_5$ and the integration constant $c_1$. (The initial value for $a$ just determines the overall scale.)   Some examples of numerical solutions for $a(\tau)$ are plotted in figure 6 for different  values for $\upsilon_5$ and $c_1$ and fixed $\kappa=1$.  In one example, $c_1=0$ and $\upsilon_5=5$, there is an initial rapid inflation followed by a linear expansion, which is very similar to the solution plotted in figure 5.  In contrast, the example,  $c_1=1$ and $\upsilon_5=0$,
a linear expansion is followed by a rapid inflation.
\begin{figure}[placement h]
\begin{center}
\includegraphics[height=3in,width=3.75in,angle=0]{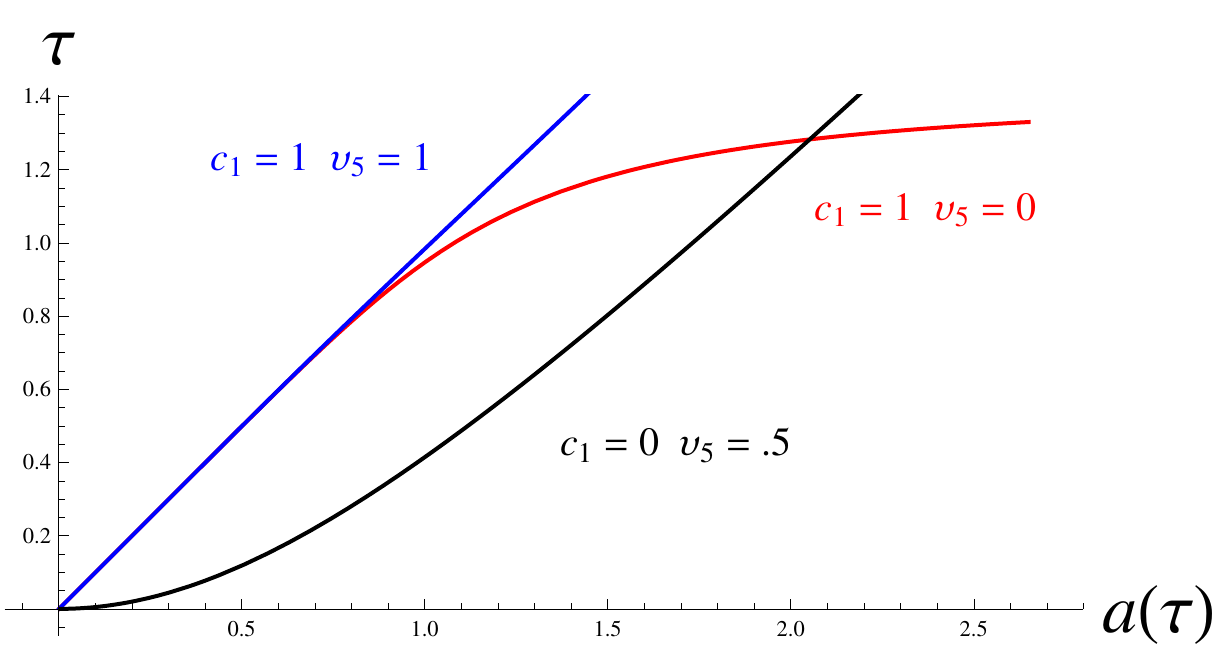}
\caption {Numerical solution for  (\ref{foftau}) and (\ref{aprmoftau}) for $\kappa=1$ and three different sets  of  values for $\upsilon_5$ and $c_1$. $\tau$ is plotted versus $a(\tau)$, with initial condition $a(0)\approx 0$ in all cases.  For the choice of $c_1=1$ and $\upsilon_5=1$, one gets an approximate linear expansion.  The choice of $c_1=0$ and $\upsilon_5=5$ has an initial rapid inflation followed by a linear expansion, while  the choice  $c_1=1$ and $\upsilon_5=0$ has
an initial linear expansion  followed by a rapid inflation.
}
\end{center}
\end{figure}
\end{enumerate}

\subsection{Perturbations about $dS^4$}

Here we consider the four-dimensional de Sitter solutions ($1.$ in subsection 5.1) which resulted upon taking the limit  $\,\omega,\,\upsilon_5 \rightarrow \infty$, with $\frac \omega{\upsilon_5}$ finite.  The solutions  have  a Poisson structure  which is determined by two finite nonvanishing parameters  $\frac \omega{\upsilon_5}$  and $\kappa$.  In this subsection we  expand about the commutative $dS^4$
solutions, expressing   perturbations in terms of commutative gauge  and scalar fields.  This requires finding the appropriate Seiberg-Witten map, which is given explicitly in Appendix C.

We again write the embedding coordinates $y^\mu$ according to (\ref{yxbta}) where $\Theta$ is the perturbation parameter and the perturbations $A^\mu$ are now functions on $dS^4$.  Locally then  $A^\mu$ are functions of $\tau,\sigma$ and  coordinates  on  $S^2$, which we take to be  the usual spherical coordinates $\theta$ and $\phi$, $0<\theta<\pi$ and $0\le \phi<2\pi$.  The action (\ref{mmactnplsqd5d})
is gauge  invariant, at least up to first order in $\Theta$.  So $A_\mu$ can  be regarded as noncommutative gauge potentials up to first order in $\Theta$.
A Seiberg-Witten map can be constructed on  $dS^4$, so that $A_\mu$ can be re-expressed in terms of commutative gauge potentials ${\cal A}_{\tt a}$, $\,{\tt a}=\tau,\sigma,\theta,\phi$, and a scalar field $\Phi$.  As in subsection 4.2, in order to produce the leading order correction to the action we need to obtain the Seiberg-Witten map up to first order.  The result is given in (\ref{ceetoo}) and (\ref{One3too})  of Appendix C.

We next  substitute the expression for $y^\mu$,  (\ref{yxbta}), along with (\ref{ceetoo}) and (\ref{One3too}), into the  action $S_c(y)$.
For the integration measure we   take
 \be d\mu_4=\frac{d\mu_{S^2}}{h(\tau,\sigma)}=\frac {\sin\theta \,d\tau d\sigma d\theta d\phi}{  h(\tau,\sigma)}=-\frac{8\kappa\upsilon_5}\omega\,\sqrt{-{\tt g}}\;d\tau d\sigma d\theta d\phi\;,\ee 
where $d\mu_{S^2}$ is the invariant measure on the sphere,
$h(\tau,\sigma)$ is given in (\ref{oneeighteen}) and ${\tt g}$ is the determinant of the metric on $dS^4$.  Thus
\be
S_c(y)\propto\int \sqrt{-{\tt g}}\;{\,d\tau d\sigma d\theta d\phi} \Bigl( -\frac {\upsilon_5 }{5}\epsilon_{\mu\nu\lambda\rho\sigma}\,y^\mu\{y^\nu, y^\lambda\}\{y^\rho, y^\sigma\}+\frac \omega 2 y_\mu y^\mu\Bigr)\label{one34}
\ee
After removing all of the total derivatives we arrive at the following simple expression for the leading order (i.e., quadratic) terms in the action $S_c(y)$ (up to a proportionality constant) 
\be
\Theta^2\int  \sqrt{-{\tt g}}\;d\tau d\sigma d\theta d\phi\, \biggl(\frac{\omega}{\upsilon_5 }\,\frac{{\cal F}_{\tau\sigma} \Phi} {a(\tau)^2\sin^2\sigma}+8\kappa^2  \frac{{\cal F}_{\theta\phi}\Phi}{\sin\theta} -12\kappa {\Phi^2}\biggr)\;,
\ee
where  $\,{\cal F}_{\tt ab}=\partial_{\tt a}{\cal A}_{\tt b}-\partial_{\tt b}{\cal A}_{\tt a}$ are the commutative field strengths.   This generalizes the effective Lagrangian in  (\ref{4twentee7}).  Just as in that example, no kinetic energy terms appear for the scalar and gauge fields, which  is not surprising since there were no kinetic energy term for $y^\mu$ in (\ref{one34}). (We therefore don't have to be concerned with the issue of the kinetic energy terms having opposing signs in this case.) 
The scalar field couples to the radial  component of the electric and magnetic fields.  The field equations imply that $\Phi$ is frozen to a constant value, while the field strengths are constrained by
\be \frac{\omega}{\upsilon_5 }\,\frac{{\cal F}_{\tau\sigma} } {a(\tau)^2\sin^2\sigma}+8\kappa^2  \frac{{\cal F}_{\theta\phi}}{\sin\theta}  =24\kappa {\Phi}\ee So for example, in the absence of an electric field, the perturbations give rise to a magnetic monopole source with charge equal to
$\;\;\frac 1{4\pi}\int {\cal F}_{\theta\phi}\,d\theta\wedge d\phi\;=\;\frac{3\Phi}{4\pi\kappa}\int{\sin\theta}\,d\theta\wedge d\phi\;=\;\frac{3\Phi}{\kappa}\; $.  The monopoles spontaneously breaks the de Sitter group symmetry down to the three-dimensional rotation group, due to the same symmetry breaking that  is present in the Poisson brackets on the surface (\ref{ntpsds4}).

\subsection{The question of rotationally invariant matrix solutions}

In subsection 2.2.1 we wrote down an ansatz for rotationally invariant matrices for the three-dimensional matrix model and their resulting equations of motion.  Here we propose to do the same in the five-dimensional case.   Rotational symmetry in this case is applied  to the three  matrices $X^i,\; i=1,2,3$ (and not also $X_4$).  This is consistent with the $SO(3)$ symmetry of the solutions in subsection 5.1 of the commutative equations of motion. 
We  show that  the  five matrix equations (\ref{eomwthqdtrm5d}) reduce  to three upon taking into  account rotational symmetry, and in a special case reduce to the matrix equations of section 3.

 A natural choice for the matrix analogues of the ans\"atse (\ref{ncprtztn5d}) and  (\ref{frdpbs}) is
\be \pmatrix{Y^0\cr Y^i\cr Y^4}= \pmatrix{t\otimes \BI\cr u \otimes j_i\cr  v\otimes \BI }\,\label{mancprtztn5d}\ee 
where $t, u, v$ are hermitean matrices and $j_i$ generate the fuzzy sphere 
\be [j_i,j_j]=i\alpha\kappa\, \epsilon_{ijk} j_k \;\ee 
Without loss of generality, we can choose its radius to be one, $j_i j_i=\BI$.  Upon substituting into the five matrix equations (\ref{eomwthqdtrm5d}) one gets that $t, u, v$ must satisfy
\beqa   -[[t,u],u]-[[t,v],v]+4i\tilde\alpha_5\alpha\kappa\;[u^2,[u,v]]_+&=&-\beta t\cr &&\cr
-[[u,t],t]+[[u,v],v]-4i\tilde\alpha_5\alpha\kappa\;[u^2,[t,v]]_++4i\tilde\alpha_5\alpha\kappa\;[[u,v],[t,u]]
+2\alpha^2\kappa^2 u^3&=&-\beta u\cr &&\cr
  -[[v,t],t]+[[v,u],u]+4i\tilde\alpha_5\alpha\kappa\;[u^2,[t,u]]_+&=&-\beta v \;, \eeqa
where $[\;,\;]_+$ denotes the anticommutator. 

While we have not found general solutions to these matrix equation, the system simplifies  when $\kappa =0$, or equivalently $\alpha=0$.   In that case the equations reduce to those of the three-dimensional system (\ref{eomwthqdtrm}) with $(t,u,v)=$ $(Y^0,Y^1,Y^2)$ and $\tilde\alpha=0$.   Therefore the recursion relations (\ref{rcrsnrltnahbne0})
can be applied in this case to numerically generate the spectra of matrices satisfying   (\ref{WAeit}).  The commutative limit of such solutions corresponds to solutions $2$ of subsection 5.1.

\subsection{Alternative background metrics}

We remark that all of the previous solutions to the commutative equations have analogs  when one changes the signature of the background metric $\eta$.
 As an example, here we consider  the  four dimensional de Sitter solution  $dS^4$ to the equations of motion (\ref{clmeaswne05dnym}).  These equations were associated with the  limit where both $\,\omega,\,\upsilon_5 \rightarrow \infty$, with $\frac \omega{\upsilon_5}$ and $\kappa$ finite and nonvanishing. If we now switch to  the Euclidean background metric  $\eta=$diag$(1,1,1,1,1)$  we recover   an $S^4$ solution to the equations of motion.   If instead we replace the background metric by    $\eta=$diag$(-1,1,1,1,-1)$  we recover  $AdS^4$.  A  fuzzy  four-sphere was obtained previously to a matrix model.\cite{Kimura:2002nq},\cite{Valtancoli:2002sm},\cite{Azuma:2004yg} However, its commutative limit differs from that of our solution.  In fact, Poisson brackets for the former don't close amongs the coordinates.   The $AdS^4$ solution  we obtain has a nontrivial Poisson structure.  The Poisson brackets are also nontrivial at the   $AdS^4$ boundary for generic cases of the constants. There is an exceptional case, however,  where the lowest order noncommutativity vanishes at the boundary, but not in the interior. It corresponds to the limit where both $\kappa$ and $\omega$ vanish. 

We first discuss the $S^4$ solution and then $AdS^4$.
\subsubsection{$S^4$}
For the  case of the Euclidean   metric   $\eta=$diag$(1,1,1,1,1)$ we can keep the  ansatz (\ref{ncprtztn5d}), only now $\tau$ is a space-like coordinate.  The equations of motion  have the solution  $ a(\tau)^2=1-\tau^2$  for $-1\le \tau\le 1$ with $ h(\tau,\sigma)$   given by (\ref{oneeighteen}).  This is the solution for $S^4$
\be  (x^0)^2+\vec x^2+(x^4)^2=1\;,\label{sfour1}\ee with Poisson brackets 
\beqa  \{x^0,x^i\}=\frac\omega{8\kappa\upsilon_5}\, \frac {x^i x^4}{( \vec x^2)^{\frac 32}}  &\qquad &
\{x^4,x^i\}=-\frac\omega{8\kappa\upsilon_5}\, \frac {x^i x^0}{( \vec x^2)^{\frac 32}} \cr  &&\cr
\{x^0,x^4\}=-\frac\omega{8\kappa\upsilon_5}\,\frac 1{\sqrt{\vec x^2}}&\qquad &
\{x^i,x^j\}=\kappa\sqrt{\vec x^2}\,\epsilon_{ijk}x^k\label{sfourpb} \eeqa
They break the $SO(5)$ rotational symmetry to  $SO(3)\times SO(2)$.
This result differs from the commutative limit of the fuzzy four-sphere.\cite{Kimura:2002nq},  \cite{Valtancoli:2002sm},\cite{Azuma:2004yg}  The algebra for the latter does not close among the embedding coordinates.  Conversely, the matrix analogue of the system (\ref{sfour1}) and (\ref{sfourpb}), if it exists, is not the fuzzy four-sphere.

\subsubsection{$AdS^4$}

A global parametrization of  $AdS^4$ is 
 \be \pmatrix{x^0\cr x^i\cr x^4}= \pmatrix{\sin{\tau}\cosh{\sigma}\cr \sinh{ \sigma}\xi^i \cr\cos{\tau}\cosh{\sigma} }\;,\label{adsonz}\ee where once again $\xi^i,\,i=1,2,3$, span the unit $(\xi^1)^2+(\xi^2)^2+(\xi^3)^2=1$.  The background     metric   is now $\eta=$diag$(-1,1,1,1,-1)$ and so
\be  -(x^0)^2+\vec x^2-(x^4)^2=-1\label{adsfrcnstrnt}\ee
The induced metric on the surface is given by
\be ds^2= -\cosh^2{\sigma} \,d\tau^2 + d\sigma^2 +\sinh^2\sigma \;d s_{{S}^2}^2\;,\ee
$d s_{{S}^2}^2$ being associated with the unit two-sphere.  The time-like parameter $\tau$ is periodic and closed time-like curves exist on this space. (\ref{adsonz}) solves the equations of motion  (\ref{clmeaswne05dnym}) upon taking the fundamental Poisson brackets to be (\ref{frdpbs}), with $h(\tau,\sigma)$ given by
\be h(\tau,\sigma)=\frac{\omega}{8\kappa\upsilon_5 }\,\frac 1{\sinh^2\sigma\cosh\sigma}\label{adsh} \ee  The resulting Poisson brackets of the coordinates $x^\mu$ of the embedding space are
\beqa  \{x^0,x^i\}=-\frac\omega{8\kappa\upsilon_5}\, \frac {x^i x^4}{( \vec x^2)^{\frac 32}}  &\qquad &
\{x^4,x^i\}=\frac\omega{8\kappa\upsilon_5}\, \frac {x^i x^0}{( \vec x^2)^{\frac 32}} \cr  &&\cr
\{x^0,x^4\}=-\frac\omega{8\kappa\upsilon_5}\,\frac 1{\sqrt{\vec x^2}}&\qquad &
\{x^i,x^j\}=\kappa\sqrt{\vec x^2}\,\epsilon_{ijk}x^k\label{gnrlpbsonads4}\eeqa
They break the $SO(3,2)$ space-time symmetry to  $SO(3)\times SO(2)$.

The result can be re-expressed in terms of  Fefferman-Graham coordinates $(z,\zeta^0,\zeta^1,\zeta^2)$, which only cover a local patch of $AdS^4$.\cite{fg}  The map from $(\tau,\sigma,\xi^i)$ to $(z,\zeta^0,\zeta^1,\zeta^2)$  is given by
$$ \zeta^0=z  \sin{\tau}\cosh{\sigma} \qquad \zeta^1= z\sinh{ \sigma}\xi^1 \qquad \zeta^2=  z\sinh{ \sigma}\xi^2$$
\be \frac 1{z} =\cos{\tau}\cosh{\sigma}-\xi^3\sinh{ \sigma} \label{zeta012}\ee
In terms of these coordinates the induced metric is given by
\be ds^2=\frac{dz^2-(d\zeta^0)^2+(d\zeta^1)^2+(d\zeta^2)^2}{z^2}\;,\ee
while the Poisson brackets are mapped to
\beqa   \{z,\zeta^0\} &=&\frac{\omega}{8\kappa\upsilon_5}\, z^3{\rm csch}\sigma\;\biggl(\xi^3{\rm ctnh}\,\sigma \cos\tau\; -\;1\biggr)
 \cr&&\cr
  \{z,\zeta^a\}&=&\kappa z^2\sinh\sigma\; \epsilon^{ab}\zeta^b -\frac{\omega}{8\kappa\upsilon_5}\,  z\,{\rm csch}^3\sigma\, \zeta^0\zeta^a
\cr  &&\cr
 \{\zeta^0,\zeta^a\}&=& \kappa z\,\sinh\sigma\; \epsilon^{ab}\zeta^0\zeta^b-\frac{\omega}{8\kappa\upsilon_5}\, z^2  {\rm csch}^3\sigma\, \zeta^a
\cr  &&\cr 
\{\zeta^1,\zeta^2\} &=&-\kappa z^3 \sinh^2\sigma  \,\Bigl(\sinh\sigma -\xi^3\cosh\sigma\cos\tau\Bigr) \;,\label{pbzetaz}\eeqa
where $a,b=1,2$. 

The boundary at infinity corresponds to $z\rightarrow 0$, or equivalently $\sigma\rightarrow\infty$ with  $ze^\sigma$ finite.  In that limit, (\ref{zeta012}) reduces to
\be \frac 2{z}e^{-\sigma}\,\zeta^0\rightarrow   \sin{\tau} \qquad\frac 2{z}e^{-\sigma}\, \zeta^a\rightarrow\xi^a  \qquad\frac 2{z}e^{-\sigma} \rightarrow \cos{\tau}-\xi^3\;\ee
These equations are solved by 
\beqa {z^2}e^{2\sigma}&\rightarrow &{(\zeta^0)^4+2(\zeta^0)^2(1-\zeta^a\zeta^a)+(1+\zeta^a\zeta^a)^2}\cr &&\cr
\cos\tau  &\rightarrow &\frac 1z e^{-\sigma}{\Bigl(1-(\zeta^0)^2+\zeta^a\zeta^a\Bigr)}\cr &&\cr
\xi^3  &\rightarrow &\frac 1z e^{-\sigma}{\Bigl(-1-(\zeta^0)^2+\zeta^a\zeta^a\Bigr)}  \label{54one}\eeqa
 We can consistently take the limit  $z\rightarrow 0$ on the Poisson brackets (\ref{pbzetaz}) since $z$ has zero Poisson brackets with the coordinates on the boundary $(\zeta^0,\zeta^1,\zeta^2)$.  The remaining Poisson brackets at $z\rightarrow 0$ are
\beqa  
 \{\zeta^0,\zeta^a\}&=&\frac 12 \kappa ze^\sigma\; \epsilon^{ab}\zeta^0\zeta^b
\cr  &&\cr 
\{\zeta^1,\zeta^2\} &=&-\frac 14\kappa z e^{\sigma}  \,\Bigl(1 +(\zeta^0)^2+\zeta^a\zeta^a \Bigr)\;, \label{five4two}\eeqa
with given   $ z e^{\sigma}$ in (\ref{54one}).
   A central element in the algebra 
is  ${\cal C} = (1-(\zeta^0)^2+\zeta^a\zeta^a)/\zeta^0$ and the Poisson bracket is then nonsingular on the two dimensional surfaces with  ${\cal C}=$constant corresponding to  symplectic leaves. 

All Poisson brackets vanish at the boundary in the limit $\kappa\rightarrow 0$.  Then for $h(\tau,\sigma)$ in  (\ref{adsh}) to be well defined, we would also need to send $\frac{\omega}{\upsilon_5}\rightarrow 0$.  Only the totally antisymmetric  term in the action (\ref{cmtvlmtscbne05d}) survives in this  case, and the equations of motion reduce to
\be \epsilon_{\mu\nu\lambda\rho\sigma}\{y^\nu,y^\lambda\}\{y^\rho, y^\sigma\} =0
\;\ee 
A general solution is 
\beqa  \{x^0,x^i\}=-\rho\, x^i x^4 &\qquad &
\{x^4,x^i\}=\rho\, x^i x^0 \cr  &&\cr
\{x^0,x^4\}=-\rho\,\vec x ^2 &\qquad &
\{x^i,x^j\}=0\;,\label{54four}\eeqa
where $\rho$ can be any function of $\vec x ^2$.  These Poisson brackets are consistent with the $AdS^4$ constraint (\ref{adsfrcnstrnt}) and the Jacobi identity.  They agree with (\ref{gnrlpbsonads4})  in the limit $\kappa,\omega\rightarrow 0$ for $\rho\sim 1/ (\vec x^2)^{\frac 32}$.
If we express $x^\mu$ using the parametrization (\ref{adsonz}), then the Poisson brackets (\ref{54four}) result from taking $\{\sigma,\tau\}=\rho\tanh\sigma$ and $\{\xi^i,\xi^j\}=0$. The Poisson brackets (\ref{54four})   generalize to any dimension $d>2$ although they may not in general solve a matrix model equation.  Once again, they can be re-expressed in terms of  Fefferman-Graham coordinates.  They vanish upon being projected to the  $AdS$ boundary $z\rightarrow 0$.  Therefore in this case the boundary is commutative (at least at lowest order), with space-time symmetry corresponding to the full three-dimensional Poincar\'e group, while the interior of $AdS $ is noncommutative.

\section{Concluding remarks}

\setcounter{equation}{0}
In the introduction we wrote down a general definition (\ref{cmxpxmwx0}) of rotationally invariant matrices embedded in three-dimensional space-time, and in  section two we  obtained  recursion relations  for such matrices which solve the  Lorentzian matrix model equations of motion.  These recursion relations allow one to generate  discrete versions of  open two-dimensional universes.   For a matrix analogue of a closed space-time solution, we need to require the existence of bottom and top states, i.e.,  there must be both a minimum and maximum time eigenvalue.  If the recursion relations are valid for such a solution, the recursion procedure must then terminate at the minimum and maximum time eigenvalues. Matrix solutions in this case would be finite dimensional.  Here and in \cite{Chaney:2015mfa}, we obtained finite-dimensional matrix solutions corresponding to  Lorentzian fuzzy spheres. 
They are  bounded  solutions to the  Lorentzian matrix model equations of motion which resolve cosmological singularities.  Here we showed that infinite-dimensional matrix solutions corresponding to the discrete series representations of the de Sitter group also resolve cosmological singularities.  In both of these examples, singularities in the induced metric  emerge after taking the continuum (or commutative) limit.    The commutative limit  also allowed for other space-times with desirable features, such as a solution which transitions from  a  rapid initial inflation to a non inflationary phase.    The quadratic term in the matrix model action studied in section three played an important role for finding novel solutions to the Lorentzian matrix model, such as the  finite dimensional  fuzzy sphere.  It was also shown to be useful for stabilizing the leading order field theory which resulted from  perturbations about the classical solutions.

Some of the matrix solutions  describing two dimensional space-times in the commutative limit  generalize in a straightforward way to higher dimensional space-time geometries, while others do not.  In section five  we saw that solutions to the commutative limit with  $\upsilon=0$ have an obvious  generalization to four dimensions. (This was case 2 in subsection 5.1.)  Since they do not require  a totally antisymmetric term in the action, analogous solutions exist in any dimension $d$. Another example of a solution which generalizes to $d>2$ is the de Sitter solution.  This is the case where the matrix model has no  kinetic energy term. In the commutative limit, the two-dimensional solution is given by (\ref{toodeeds}) and (\ref{cmdS2alg}), while its four-dimensional counterpart (case 1 in subsection 5.1) is given by (\ref{onenineone}) and (\ref{ntpsds4}).   The corresponding $S^4$  and $AdS^4$ solutions along with their attached Poisson structures were given explicitly in subsection 5.4.  The Possion brackets of the $S^4$ solution differed from the
commutative limit of the  fuzzy  four-sphere  obtained in \cite{Kimura:2002nq},\cite{Valtancoli:2002sm},\cite{Azuma:2004yg}. This is obvious because the  commutation algebra for the fuzzy four-sphere  don't close. On the other hand, due to the nontrivial nature of our Poisson brackets for the $d=4$ commutative solutions, the matrix model analogues of these  solutions are not straightforward to obtain, unlike the case with $d=2$.    Concerning our $AdS^4$ solution, we found that the  general Poisson brackets (\ref{gnrlpbsonads4})  are nonzero when projected to  the  boundary.   The exceptional case corresponds to the limit where both $\kappa$ and $\omega$ vanish.  In this case the boundary remains commutative, at least at lowest order, where the space-time symmetry  is the full three-dimensional Poincar\'e group.    The  corresponding matrix action in this case consists only  of the totally antisymmetric term,  $\epsilon_{\mu\nu\lambda\rho\sigma}{\rm Tr}Y^\mu Y^\nu Y^\lambda Y^\rho Y^\sigma$.  These  results could  have interesting implications for the $AdS/CFT$ correspondence.

 In addition to the solutions which generalize from $d=2$, there are some solutions to the higher dimensional theories which have no $d=2$ analogue.  This was true for case 3 in subsection 5.1.
Concerning the different  families of $d=4$ space-time  manifolds  obtained in subsection 5.1, it may be possible to find other solutions to the commutative equations, and even the  matrix equations.  For example, we can use the fact that fuzzy coset models are  higher dimensional  generalizations  of the the fuzzy  sphere.\cite{Balachandran:2005ew} The latter was shown in \cite{Chaney:2015mfa}   to solve the three-dimensional Lorentzian matrix model, and so it is natural to ask if the former solve higher dimensional Lorentzian matrix models.  More specifically, fuzzy $CP^2$ may solve the  five-dimensional model. In the commutative limit, the solutions should yield cosets manifolds embedded in Minkowski space-time.  Another possibility for finding more solutions is to modify the ansatz (\ref{ncprtztn5d}) in section five, which for any time slice describes $S^3$.   For example, we can let the spatial coordinates $x^1,\;x^2,\;x^3,\;x^4$ instead span $S^2\times S^1$.  For this we can introduce a second radial quantity $b(\tau)$ and replace (\ref{ncprtztn5d})  by
\be \pmatrix{x^0\cr x^i\cr x^4}= \pmatrix{\tau\cr \Bigl(a( \tau) \sin\sigma+b(\tau)\Bigl)\xi^i\cr a( \tau)\cos\sigma}\ee  
  Here $\sigma$ is a periodic parameter, $0\le \sigma < 2 \pi$ .  Its canonical conjugate will have a regularly spaced spectrum in the noncommutative version of the theory, similar to that of the operator $\hat t$ in (\ref{tmspctrm}).  This ansatz is a generalization of (\ref{ncprtztn5d}), since it reduces to it  in the limit $b\rightarrow 0$.   $SO(3)$ is an isometry for this system, and this three dimensional rotation symmetry is preserved if we once again impose the Poisson brackets  (\ref{frdpbs}).

While the focus of this article has been to search for matrix model solutions which give rise to cosmological space-times in the continuum limit, one can have for matrix analogues of other solutions for general relativity, such as black hole solutions.\cite{Blaschke:2010ye} The  eigenvalues of such a matrix solution gives a lattice description of a black hole. Bounded solutions would necessarily give a resolution of the black hole singularity.  It would be of interest to demonstrate how to recover black hole metrics, along with their singularities, from the induced metric upon taking the continuum limit.  

We examined perturbations about the rotationally invariant solutions to the three-dimensional Lorentzian matrix in section four, and the  de Sitter solution of the five-dimensional model  in  subsection 5.2.  In the commutative limit the result is a gauge theory coupled to a scalar field theory.  The gauge fields are associated with longitudinal perturbations, while the scalar fields denote perturbations normal to the space-time surface.  A persistent feature of the emergent field theory is that the kinetic energies of the gauge  and scalar fields have opposite signs.  This presents no obstacle to the two-dimensional field theories, since the gauge fields are nondynamical and can be eliminated from the action.  The result is an effective field theory for the remaining scalar field which for different choices of the parameters can be massive, massless or tachyionic.  It was also not an issue for perturbations about the four-dimensional de Sitter solution, as the kinetic energy vanished in that case.  This system led to  magnetic monopoles on the surface.  For more general matrix solutions leading to space-time manifolds with dimension greater than two, the difference in signs in the kinetic energy terms remains an issue which requires a creative solution.

\medskip

\begin{appendices}

\section{ 2D Rotationally invariant Seiberg-Witten map}
\setcounter{equation}{0}
\renewcommand{\theequation}{A.\arabic{equation}}

Here we review the general Seiberg-Witten map  up to first order
in $\Theta$ on a two dimensional rotationally invariant surface.\cite{Stern:2014aqa} It is required to be consistent with (\ref{pbtauphi}) and (\ref{ncprtztn}).
 At lowest order in $\Theta$, contributions to the noncommutative potentials $A_\mu$ come from the commutative gauge potentials $({\cal A}_\tau,{\cal A}_\sigma)$ along the tangent directions to the surface, and the scalar field is associated with perturbations normal to the surface.  Also at lowest order, the noncommutative gauge parameter $ \Lambda$ can be identified with the commutative gauge parameter $\lambda$. The next order is obtained by demanding consistency with (\ref{gvrtnsAmu}). The result is
\beqa A_\mu&=& A^{(0)}_\mu+\Theta A^{(1)}_\mu+{\cal O}(\Theta^2)\cr&&\cr
\Lambda&=&\Lambda^{(0)}+\Theta \Lambda^{(1)}+{\cal O}(\Theta^2)\;,\label{axpnnbta}\eeqa
\beqa A^{(0)0}&=&- h(\tau)\Bigl( -{\cal A}_\sigma+ {a'(\tau)}a(\tau)\phi\Bigr)\cr&&\cr A^{(0)}_\pm &=&-h(\tau) e^{\pm i\sigma}\Bigl(\pm ia(\tau){\cal A}_\tau- a'(\tau){\cal A}_\sigma+a(\tau)\phi\Bigr) \cr&&\cr \Lambda^{(0)}&=&\lambda\label{swzero}\;
\eeqa
\beqa A^{(1)0}&=& h(\tau)\biggl(\frac 1 2 \partial_\tau \Bigl( h(\tau){\cal A}_\sigma^2\Bigr)+a'(\tau)h(\tau) {\cal A}_\tau\partial_\sigma\Bigl(a(\tau)\phi\Bigr)-{\cal A}_\sigma\partial_\tau\Bigr( a'(\tau)h(\tau)a(\tau)\phi\Bigl)\biggr)
\cr&&\cr A^{(1)}_\pm &=&h(\tau) e^{\pm i\sigma}\biggl( \mp i \partial_\tau\Bigr( a(\tau)h(\tau) {\cal A}_\tau\Bigr){\cal A}_\sigma\mp i a(\tau)h(\tau) {\cal A}_\tau {\cal F}_{\tau\sigma}\pm i h(\tau) a(\tau) {\cal A}_\tau\phi\cr&&\cr &&\quad+ h(\tau) {\cal A}_\tau \partial_\sigma \Bigl(a(\tau)\phi\Bigr)-{\cal A}_\sigma \partial_\tau\Bigl(h(\tau)a(\tau)\phi\Bigr)+\frac 12 \partial_\tau\Bigr( a'(\tau)h(\tau) {\cal A}_\sigma^2 \Bigr)-\frac12 a(\tau)h(\tau){\cal A}_\tau^2\biggr)
\cr&&\cr \Lambda^{(1)}&=& h(\tau) {\cal A}_\sigma\partial_\tau\lambda\label{swone}
\eeqa
where $A^{(n)}_\pm = A^{(n)}_1\pm A^{(n)}_2$ and
${\cal F}_{\tau\sigma}=\partial_\tau {\cal A}_\sigma-\partial_\sigma {\cal A}_\tau$ is the $U(1)$ gauge field on the surface.

\section{$\;\;$Seiberg-Witten map on $dS^4$}
\setcounter{equation}{0}
\renewcommand{\theequation}{B.\arabic{equation}}

Here we obtain the consider the Seiberg-Witten map up to first order in the noncommutativity parameter for the four-dimensional de Sitter solution of section 5.

We first obtain the zeroth order result.
This  is easy to determine by comparing the gauge transformation properties of  the commutative gauge potentials ${\cal A}_{\tt a}$, $\,{\tt a}=\tau,\sigma,\theta,\phi$ with those of the noncommutative potentials $A_\mu$, using the Poisson brackets (\ref{frdpbs}).  The gauge variations of the former are simply $\delta{\cal A}_{\tt a}=\partial _{\tt a}\lambda$, $\lambda$ being an infinitesimal commutative gauge parameter on $dS^4$, while the latter is given by (\ref{gvrtnsAmu}), where $\Lambda$ is an   infinitesimal noncommutative gauge parameter.  The result is 
\be\delta A_\mu=- h  \,\Bigl(\partial_\tau\Lambda\partial_\sigma x_\mu-\partial_\sigma\Lambda\partial_\tau x_\mu\Bigr)+\frac \kappa{\sin\theta}\Bigl(\partial_\theta\Lambda\partial_\phi x_\mu-\partial_\phi\Lambda\partial_\theta x_\mu\Bigr)+\Theta \delta A^{(1)}_\mu+{\cal O}(\Theta^2)\;,
\ee
where $h=h(\tau,\sigma)$ is given in (\ref{oneeighteen}).  
 Then at  zeroth order in $\Theta$ the commutative gauge potentials are tangent to $dS^4$, while an additional degree of freedom $\Phi$ is associated with  perturbations normal to the surface.  Thus the zeroth order result  for $A_\mu$ and $\Lambda$ is given in
\beqa A_\mu&=&- h \,\Bigl({\cal A}_\tau\partial_\sigma x_\mu-{\cal A}_\sigma\partial_\tau x_\mu\Bigr)+\frac \kappa{\sin\theta}\Bigl({\cal A}_\theta\partial_\phi x_\mu-{\cal A}_\phi\partial_\theta x_\mu\Bigr)+ \Phi\, x_\mu+\Theta A^{(1)}_\mu+{\cal O}(\Theta^2)\cr &&\cr \Lambda &=&\lambda + \Theta \Lambda^{(1)}+{\cal O}(\Theta^2) \label{ceetoo}\eeqa

For the first order terms, $A^{(1)}_\mu$  and $ \Lambda^{(1)}$, we demand consistency with  (\ref{gvrtnsAmu}).  A solution is 
\beqa  
A^{(1)}_\mu 
&=&\partial_\tau x_\mu \,\Biggl(
\frac{a^3 a'}2\biggl( h^2{\cal A}_\tau^2+\kappa^2\sin^2\sigma {\cal A}_\Omega^2\biggr)+\frac{ h a}2\partial_\tau\Bigl(
\frac h a {\cal A}_\sigma^2\Bigr)-\frac{ h\kappa}{\sin\theta} \Bigl({\cal A}_\theta {\cal F}_{\sigma\phi}+{\cal A}_\phi\partial_\theta{\cal A}_\sigma\Bigr)
\Biggr)\cr 
&&\cr 
&+&{\partial_\sigma x_\mu }\Biggl( h^2{\cal A}_\tau {\cal F}_{\sigma\tau}-\frac ha{\cal A}_\sigma\partial_\tau(a h {\cal A}_\tau)+\frac 14 \partial_\sigma h^2 {\cal A}_\tau^2-\frac{\kappa^2\sin(2\sigma)} 4{\cal A}_\Omega^2+\frac{ h\kappa}{\sin\theta} \Bigl({\cal A}_\theta {\cal F}_{\tau\phi}+{\cal A}_\phi\partial_\theta{\cal A}_\tau\Bigr)\Biggr)\cr 
&&\cr 
&+&\,\frac{\kappa\,\partial_\theta x_\mu }{\sin\theta}\,\Biggl({ h }\Bigl({\cal A}_\tau{\cal F}_{\sigma\phi}+\cot\sigma{\cal A}_\tau{\cal A}_\phi-\frac 1 a{\cal A}_\sigma\partial_\tau(a{\cal A}_\phi) \Bigr)+\frac{\kappa}{2}
\Bigl( \frac {\partial_\theta    {\cal A}_\phi^2 }{\sin\theta}-  \cos\theta {\cal A}_\Omega^2\Bigr)
\Biggr)
\cr &&\cr 
&-&\,\frac{\kappa\,\partial_\phi x_\mu}{\sin\theta} \,\Biggl( h\,\Bigl({\cal A}_\tau{\cal F}_{\sigma\theta}+ \cot\sigma {\cal A}_\tau{\cal A}_\theta-\frac 1 a{\cal A}_\sigma \partial_\tau(a {\cal A}_\theta) \Bigr)+\frac{ \kappa }{\sin\theta}\Bigl( {\cal A}_\theta {\cal F}_{\theta\phi}+{\cal A}_\phi\partial_\theta{\cal A}_\theta\Bigr)\Biggr)
\cr&&\cr
&-&  \,\frac  {x_\mu }2 \,\Biggl( h^2 \Bigl(a^2 {\cal A}_\tau^2 -\frac {{\cal A}_\sigma^2}{a^2}\Bigr)+\kappa^2 a^2\sin^2\sigma{\cal A}_\Omega^2\Biggr)
\cr &&\cr 
&-& \,h \,\biggl({\cal A}_\tau\partial_\sigma (\Phi x_\mu) -{\cal A}_\sigma\partial_\tau (\Phi x_\mu)\biggr)+\frac \kappa{\sin\theta}\biggl({\cal A}_\theta\partial_\phi (\Phi x_\mu)-{\cal A}_\phi\partial_\theta (\Phi x_\mu)\biggr)
\cr&&\cr &&\cr 
\Lambda^{(1)}&=& h {\cal A}_\sigma\partial_\tau\lambda  -\frac \kappa{\sin\theta}{\cal A}_\phi\partial_\theta\lambda\;,
\cr &&\label{One3too}
\eeqa
where we define ${\cal A}_\Omega^2={\cal A}_\theta^2+{{\cal A}_\phi^2}/{\sin^2\theta }$. $\,{\cal F}_{\tt ab}=\partial_{\tt a}{\cal A}_{\tt b}-\partial_{\tt b}{\cal A}_{\tt a}$ are the commutative field strengths.  In obtaining (\ref{One3too})  we have used the explicit expression for the de Sitter solution, $a^2=\tau^2+1$.

\end{appendices}

\bigskip
{\Large {\bf Acknowledgments} }

\noindent
We are very grateful to A. Pinzul and C.  Uhlemann for valuable discussions.

\end{document}